\documentclass[a4paper,11pt,colorlinks=true,urlcolor=blue,anchorcolor=blue,citecolor=blue,filecolor=blue,linkcolor=blue,menucolor=blue,linktocpage=true,pdfa=true,allcolors=black]{article}
\usepackage{jinstpub} 
\usepackage{lineno}
\RequirePackage{fix-cm}

\RequirePackage{mathptmx}      
\RequirePackage{graphicx}
\usepackage{rotating}

\RequirePackage{amssymb}
\RequirePackage{multirow}
\usepackage[dvipsnames]{xcolor}
\RequirePackage{amsmath}
\RequirePackage{lineno}
\RequirePackage{booktabs}
\RequirePackage{float}
\RequirePackage{comment}
\RequirePackage[version=4]{mhchem}
\RequirePackage{booktabs}
\RequirePackage{url}

\RequirePackage[multiple]{footmisc}
\RequirePackage{upgreek}
\RequirePackage{mathtools}
\RequirePackage{cuted}
\RequirePackage{relsize}
\RequirePackage{makecell}
\RequirePackage{tablefootnote}
\RequirePackage{siunitx}
\usepackage[acronym]{glossaries}

\newcommand{\CC}{C\nolinebreak\hspace{-.05em}\raisebox{.4ex}{\tiny\bf +}\nolinebreak\hspace{-.10em}\raisebox{.4ex}{\tiny\bf +}}
\def\CC{{C\nolinebreak[4]\hspace{-.05em}\raisebox{.4ex}{\tiny\bf ++}}}

\newcommand{\opticalplanes}{{\SI{21}{\square \meter}}}
\newcommand{\TPCnchan}{{2112}}
\newcommand{\IVnchan}{{480}}
\newcommand{\OVnchan}{{128}}

\newcommand{\pdudim}{20$\times$20 \SI{}{\square \centi \meter}}

\newcommand{\uarvolume}{{\SI{50}{tonnes}}}
\newcommand{\ivvolume}{{\SI{36}{tonnes}}}
\newcommand{\ovvolume}{{\SI{650}{tonnes}}}

\newcommand{\dsheight}{\SI{348}{\centi\meter}}
\newcommand{\dsdiameter}{\SI{350}{\centi\meter}}
\newcommand{\SPEamplitude}{\SI{14}{\milli\volt}}
\newcommand{\SPESNR}{6}
\newcommand{\SignalRisetime}{\SI{100}{\nano\second}}
\newcommand{\SignalDecaytime}{\SI{300}{\nano\second}}

\newcommand{\timeslice}{{\SI{1} {\second}}}
\newcommand{\drifttime}{{\SI{5}{\milli \second}}}

\newcommand{\digmodel}{VX2745}

\newcommand{\wfds}{WFDs}

\newcommand{\controlswitch}{S2805S-24TF}

\makeglossaries
\newacronym{daq}{DAQ}{Data Acquisition}
\newacronym{tpc}{TPC}{Time Projection Chamber}
\newacronym{snr}{SNR}{Signal-to-Noise Ratio}
\newacronym{fep}{FEP}{Front End Processor}
\newacronym{tsp}{TSP}{Time Slice Processor}
\newacronym{wfd}{WFD}{Waveform digitiser}
\newacronym{pdu}{PDU}{Photon Detection Unit}
\newacronym{pm}{PM}{Pool Manager}
\newacronym{gdm}{GDM}{Global Data Manager}
\newacronym{cdm}{CDM}{Crate Data Manager}
\newacronym{ds20k}{DS-20k}{DarkSide-20k}
\newacronym{dm}{DM}{Dark Matter}
\newacronym{wimp}{WIMP}{Weakly Interacting Massive Particles}
\newacronym{sipm}{SiPM}{Silicon PhotoMultiplier}
\newacronym{iv}{IV}{Inner Veto}
\newacronym{ov}{OV}{Outer Veto}
\newacronym{tpb}{TPB}{TetraPhenyl Butadiene}
\newacronym{cdu}{CDU}{Cabinet Distribution Unit}
\newacronym{scp}{SCP}{Slow Control Processor}
\newacronym{midas}{MIDAS}{Maximum Integrated Data Acquisition System}
\newacronym{dcs}{DCS}{Detector Control System}
\newacronym{dsp}{DSP}{Digital Signal Processor}
\newacronym{cnaf}{CNAF}{Centro Nazionale Analisi Fotogrammi}
\newacronym{sc}{SC}{Slow Control}
\newacronym{uar}{UAr}{Underground Argon}
\newacronym{aar}{AAr}{Atmospheric Argon}
\newacronym{hv}{HV}{High Voltage}
\newacronym{lv}{LV}{Low Voltage}
\newacronym{ts}{TS}{Time Slice}
\newacronym{tsm}{TSM}{Time Slice Marker}
\newacronym{lut}{LUT}{Lookup Table}
\newacronym{fifo}{FIFO}{First-In First-Out}
\newacronym{lngs}{LNGS}{Laboratori Nazionali del Gran Sasso}
\newacronym{fir}{FIR}{Finite Impulse Response}
\newacronym{ps}{PS}{Power Supply}
\newacronym{zle}{ZLE}{Zero Length Encoding}

\begin{document}

\title{The Darkside-20k Data Acquisition System}








\author[1]{Fabio Acerbi}
\author[2]{Pushparaj Adhikari}
\author[3,4]{Paolo Agnes}
\author[5]{Iftikhar Ahmad}
\author[6,7]{Sebastiano Albergo}
\author[8]{Ivone F.M. Albuquerque}
\author[9]{Thomas Olling Alexander}
\author[10]{Andrew Knight Alton}
\author[11]{Pierre-Andr\'e Amaudruz}
\author[6,7]{Gioacchino Alex Anastasi}
% ...

\affiliation[1]{Fondazione Bruno Kessler, Povo, 38123, Italy}
\affiliation[2]{Department of Physics, Carleton University, Ottawa, ON K1S 5B6, Canada}
\affiliation[3]{Gran Sasso Science Institute, L'Aquila, 67100, Italy}
\affiliation[4]{INFN Laboratori Nazionali del Gran Sasso, Assergi (AQ), 67100, Italy}
\affiliation[5]{Center for Experimental Cosmology \& Instrumentation,, Riverside, CA 92507, USA}
\affiliation[6]{INFN Catania, Catania, 95121, Italy}
\affiliation[7]{Universit\`a of Catania, Catania, 95124, Italy}
\affiliation[8]{Instituto de F\'isica, Universidade de S\~ao Paulo, S\~ao Paulo, 05508-090, Brazil}
\affiliation[9]{Pacific Northwest National Laboratory, Richland, WA 99352, USA}
\affiliation[10]{Physics Department, Augustana University, Sioux Falls, SD 57197, USA}
\affiliation[11]{TRIUMF, 4004 Wesbrook Mall, Vancouver, BC V6T 2A3, Canada}
% ...

\emailAdd{ds-ed@lists.infn.it}

\abstract{DarkSide-20k  is a \gls{wimp} search experiment using liquid argon as a target, designed to perform a background-free search for dark matter with unprecedented sensitivity, and is currently under construction at INFN Laboratori Nazionali del Gran Sasso, Italy. The detector comprises a dual-phase Time Projection Chamber complemented with external veto systems and is equipped with a total of 2720 \gls{sipm}-based readout channels. This work presents the \gls{daq} system designed for DarkSide-20k. The system is capable of continuous, triggerless digitisation of the waveforms with high single-photoelectron detection efficiency and online processing, ensuring data reduction for long-term storage.

The DarkSide-20k \gls{daq} system employs commercial CAEN VX2745 digitisers with custom FPGA firmware implementation, identifying pulses in the digitised waveforms. Timing and synchronisation across all 48 digitisers are provided by custom Global and Crate Data Manager boards distributing a phase-aligned clock derived from a disciplined rubidium standard. Waveform segments are processed in real time by Front End Processor machines. Data are organised into collections containing whole detector information and distributed across a farm of Time Slice Processors for event reconstruction, classification, and further reduction before storage and offline analysis.

A full “Quadrant” of the system, corresponding to one quarter of the final DAQ, has been assembled and validated at TRIUMF laboratory in Canada. The Quadrant has been stress-tested with simultaneous pulses and demonstrated sustained digitiser readout exceeding expected physics rates and stable long-term performance.}



\maketitle
\flushbottom

\section{Introduction}
\label{sec:intro}
This paper describes the \gls{daq} system of the \gls{ds20k} experiment.
\gls{ds20k} is a next-generation, multi-tonne \gls{dm} detector under construction at the INFN \gls{lngs}. Designed to achieve leading sensitivity in the search for \glspl{wimp} over the next decade, \gls{ds20k} will probe the \gls{dm} mass range from 1 GeV/$\mathrm{c}^2$ to 10 TeV/$\mathrm{c}^2$ via Nuclear Recoil (NR). The experiment is designed to reach the sensitivity level where solar and atmospheric neutrinos become significant backgrounds~\cite{tdr,ds20klight}.

\begin{figure}[tpb]
    \centering
    \includegraphics[width=0.9\linewidth]{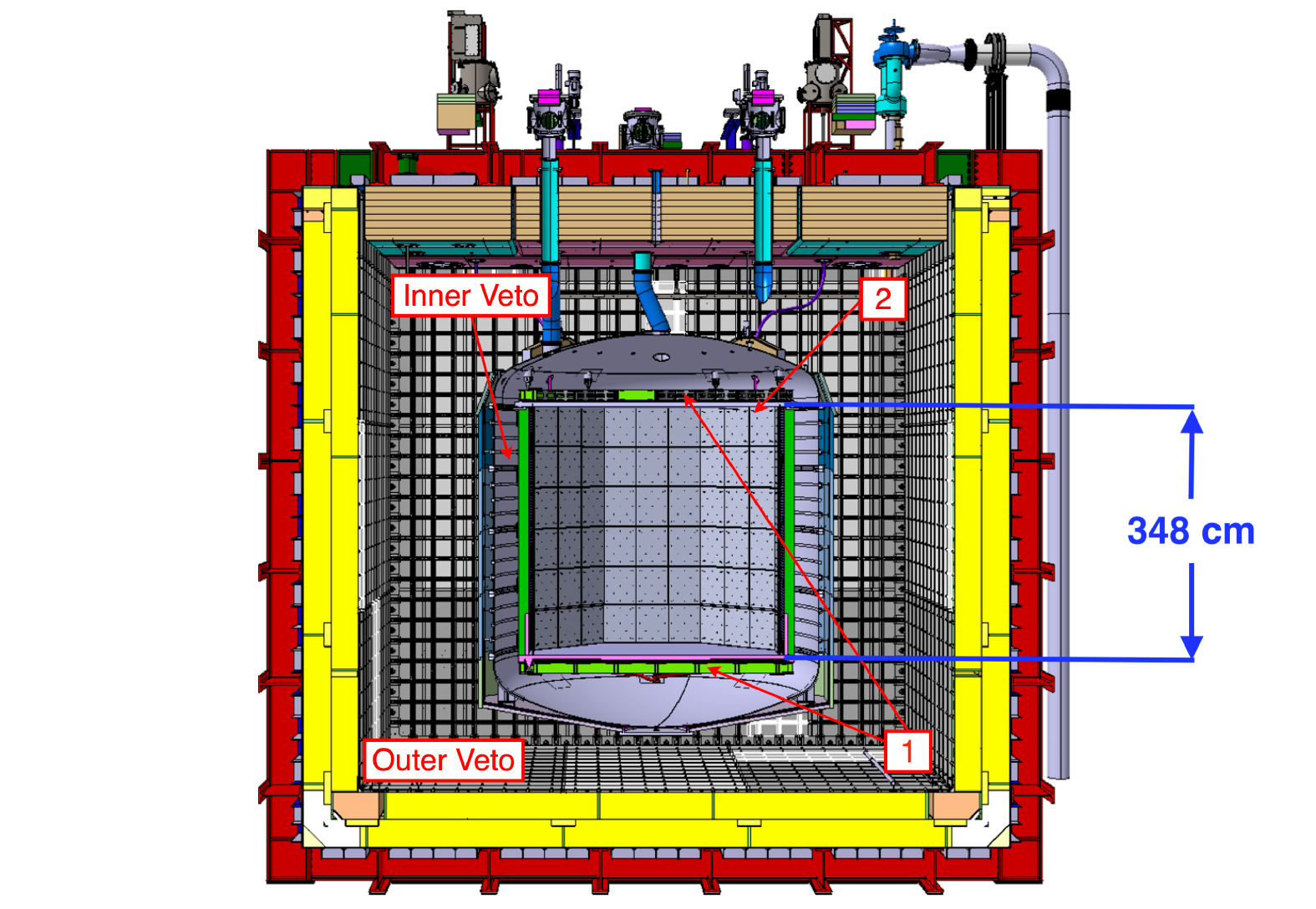}
    \caption{Cross-sectional view of the \gls{ds20k} detector. The \gls{tpc} is equipped with top and bottom Optical Planes (1). At the top of the \gls{tpc} (2), a thin layer of gaseous argon together with a stainless-steel grid enables the production of the electroluminescence signal. The \gls{tpc} is enclosed in the Inner and Outer Vetos (IV and OV).}
    \label{fig:DS-detector}
\end{figure}

The core detection system of \gls{ds20k} is a dual-phase \gls{tpc} with a vertical electron drift field (see ~\autoref{fig:DS-detector}). The \gls{tpc} is a \dsheight\ tall octagonal prism made of transparent acrylic with a \dsdiameter\ inner diameter, containing \uarvolume\ of liquid argon extracted from underground sources, that acts as a \gls{dm} target~\cite{tdr}.

The \gls{tpc} is instrumented with two arrays of \gls{sipm}-based photosensors, organised into \TPCnchan\ readout channels.
These arrays, called Optical Planes, are placed on the top and bottom of the \gls{tpc} and provide an optical coverage of \opticalplanes.
\glspl{sipm} are assembled in \(49.5\) $\times$ \(49.5\) \SI{}{\milli\meter\squared} modules referred to as \textit{Tiles}.  
Signals from the Tiles are read out by cryogenic transimpedance amplifiers, whose feedback network sets the gain and bandwidth of the single-photoelectron response. On the PDU motherboard, the single-ended outputs of four Tiles are actively summed in an analogue summing stage, further amplified, converted to differential signals, and delivered to the Waveform Digitisers (\wfds) operating at room temperature outside the cryostat.   A readout channel is made by the analogue sum of the signals from four Tiles~\cite{Razeto2022,acerbi2024qualityassurancequalitycontrol,DarkSide-20k:2025avf_QAQCtiles}. Four channels (i.e. 16 Tiles) are hosted on a mechanically independent unit of size \pdudim\ known as \glspl{pdu}.
Single photo-electron signals from one channel exhibit, at the digitiser level and under typical operating conditions, a \SPEamplitude\ amplitude with a signal to noise ratio of about \SPESNR, a rise time of approximately \SignalRisetime, and an exponential decay of about \SignalDecaytime.

The \gls{tpc} Optical Planes detect the prompt scintillation signal (S1) produced by an interaction in liquid argon and  the secondary signal  (S2) from electroluminescence by the ionisation electrons as they pass through the gas layer below the anode~\cite{Agnes:2015gu}.

Electric fields inside the \gls{tpc} are defined by three electrodes: the cathode on the inner face of the bottom cap of the \gls{tpc}, the anode on the inner face of the top cap of the \gls{tpc}, and a grid of wires covering the cross section of the \gls{tpc} positioned \SI{1}{\centi\meter} below the anode. During standard operation, the grid is immersed \SI{3}{\milli\meter} below the liquid argon surface. Above the surface, a \SI{7}{\milli\meter} thick layer of gaseous argon separates the liquid from the anode. The drift field (typically \SI{200}{\volt \per \centi\meter}) is established between the cathode and the wire-grid and is responsible for transporting the ionization electrons produced by particle interactions in the liquid argon toward the liquid–gas interface. The extraction field (typically \SI{5}{\kilo\volt \per \centi\meter}), established between the wire-grid and the anode, enables the efficient extraction of these electrons from the liquid into the gas phase and accelerates them toward the anode, where they generate proportional scintillation light. Graded field rings are used on the inner walls of the \gls{tpc} to ensure a uniform field in the drift volume. The cathode, anode and field ring electrodes are made using Clevios\textsuperscript{\texttrademark}, a transparent conductive coating. An arrangement of high-efficiency reflectors cover the inner walls of the \gls{tpc}. The reflectors, the anode and the cathode are coated with \gls{tpb} wavelength shifter to convert the argon scintillation (in the VUV)  to the sensitive range of the SiPMs~\cite{Benson_2018}.

The \gls{tpc} is surrounded by a stainless steel vessel containing an additional \ivvolume\ of underground argon, forming the \gls{iv}. This vessel is further immersed in \ovvolume\ of liquefied atmospheric argon within a DUNE-like membrane cryostat, serving as the \gls{ov}~\cite{protodune,Abi_2020}. Both veto systems are instrumented with \gls{sipm}-based photosensors to collect the scintillation light. Sensors are arranged in veto-PDU units (vPDU) located on the external surfaces of the \gls{tpc} and on the external surface of the stainless steel vessel. The vPDU differs from the PDU only in the amplification stage, provided by a custom asic chip that delivers similar gain, signal shape and signal-to-noise ratio than for \gls{tpc} tiles,  for a total of
\IVnchan\ and \OVnchan\ channels in the \gls{iv} and  \gls{ov}, respectively~\cite{Kugathasan:2020xry,DarkSide-20k:2026cdf_QAQCvTiles}. 
The \gls{iv} enables efficient neutron tagging, which is critical since neutrons can mimic \gls{dm} interactions in the \gls{tpc}. The \gls{ov} provides additional passive shield against external neutrons and acts as an active cosmic muon veto. 

\subsection{The DarkSide-20k DAQ}
\label{sec:ds20k-daq}
The \gls{ds20k} \gls{daq} is designed to continuously acquire signals from the \gls{tpc} and Veto photosensors. Analogue waveforms are digitised and transferred to the next stage for processing without waiting for a trigger decision. In this sense, the \gls{ds20k} experiment operates in \textit{triggerless mode}, where the data stream is uninterrupted, and the isolation of interesting signals for physics searches is offloaded to an online computing farm (\autoref{sec:time_slice_architecture}). This architecture is engineered to avoid biases from 
any specific trigger configuration and decisions based on incomplete detector information, enabling the search for a wide range of dark matter and astrophysical signals, like supernova neutrino bursts.
The design of the \gls{daq} system must satisfy stringent performance requirements: (i) ensure a sensitivity to single photoelectrons with an efficiency greater than 90\%; (ii) handle an input rate of approximately 100 physical interactions per second in the \gls{tpc}, increasing up to 200 per second during calibration runs, where each interaction may produce an S1 signal, an S2 signal, or an S1--S2 pair (i.e an \textit{event}); and (iii) reduce the data volume from about \SI{3} GB/s at the digitiser level to the projected \SI{60} MB/s on permanent storage during standard operations, and below \SI{200} MB/s during calibration.

The expected event rate in the Inner Detector (\gls{tpc} and \gls{iv}) is of the order of 200 events per second. The S1 signals in the energy range relevant for \gls{dm} searches consist of about 100 photo-electrons spread rather uniformly over the photo-sensor planes. Because of the large number of readout channels, the signal of interest consists mainly of single photo-electrons. This allows for a significant data reduction if parameters such as pulse charge, timing, and prominence (peak amplitude over the baseline) can be computed online. At higher energies, S1 signals are still characterised by relatively short pulses, thus posing little burden to the \gls{daq} system.  

The typical S2 signal is of the order of thousands of PEs, a factor of 10, or more, larger than its accompanying S1. The primary challenge for the  \gls{daq} system arises from the high data rate generated by the S2 light in the \gls{tpc}. Typically, only 25\% of the light is concentrated in a 3x3 readout channel matrix immediately above the position of the interaction in the $x-y$ plane. The remaining light is sparse over all the \gls{tpc} channels, including the bottom plane. Due to substantial amplification of the S2 signal in the gas phase a high number of channels need to be read with long acquisition windows simultaneously. This constitutes a challenge for the data acquisition. The high event rate expected in the \gls{ov} requires the implementation of less stringent requirements in terms of single photo-electron efficiency and timing resolution, the details of which are still under development.


\section{DAQ architecture}
\label{sec:daq_network}

\begin{figure*}[tpb]
    \centering
    \begin{minipage}{0.93\textwidth}
            \begin{minipage}{\textwidth}
                \centering
                \includegraphics[width=0.9\textwidth]{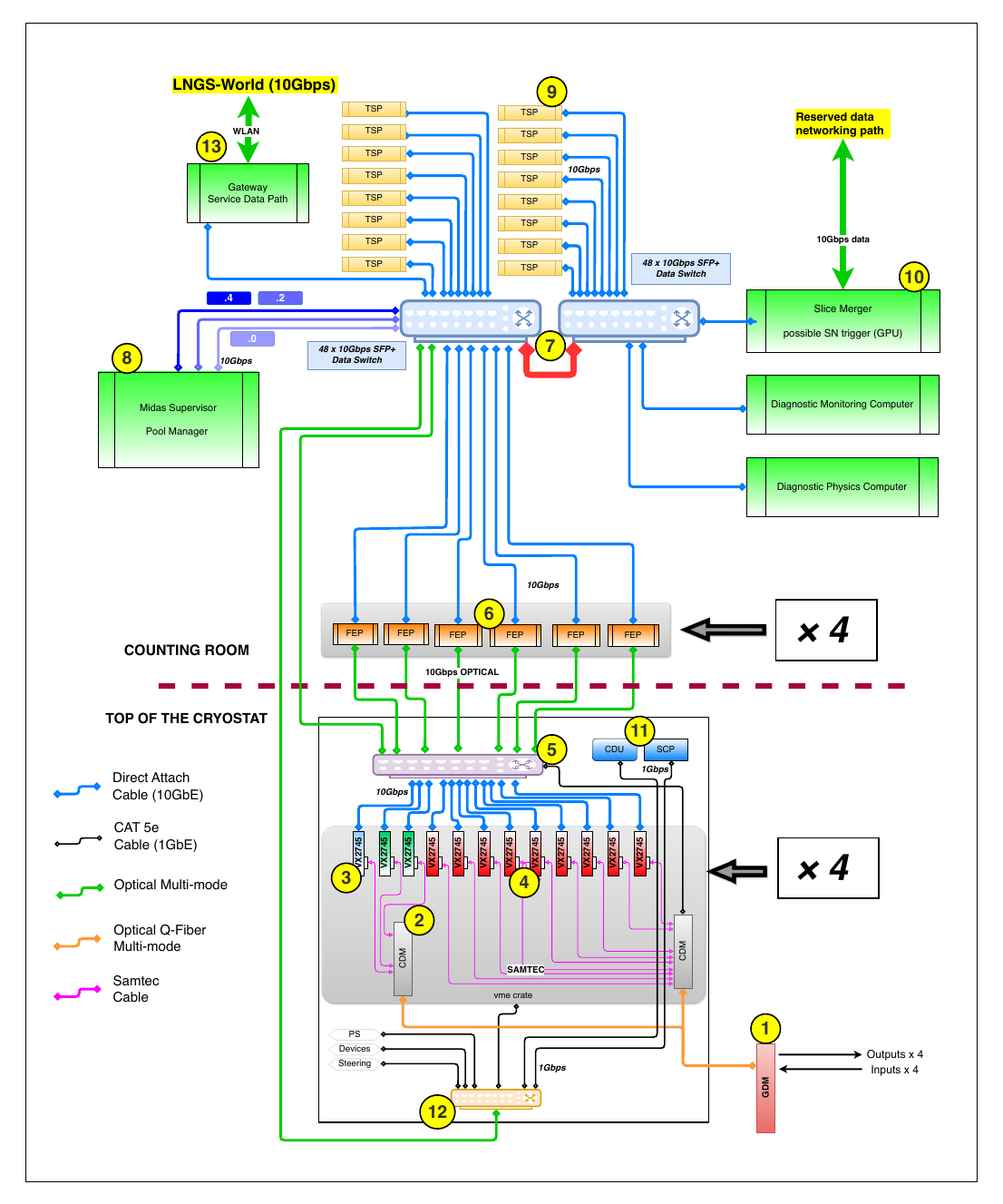}
                \caption{Schematic representation of the Data Acquisition (DAQ) network. The Global Data Manager (GDM, \(1\)) and the Crate Data Managers (CDMs, \(2\)) distribute clock and command signals to the Waveform Digitisers (WFDs) serving the veto systems and the Time Projection Chamber (TPC, \(3\) and \(4\), respectively). Each WFD is connected through a network switch (\(5\)) to dedicated Front End Processor computers (FEPs, \(6\)), where digitised waveforms are processed. Optical \SI{10} GbE links connect the WFD switches to the FEPs, in turn connected to the main data switch (\(7\)). The DAQ server, hosting the Pool Manager (PM, \(8\)), distributes detector data to a cluster of Time Slice Processors (TSPs, \(9\)), where online reconstruction is performed. Data from multiple TSPs are collected by the Merger machine (\(10\)) and stored on local disk before transfer through the Gateway machine (\(13\)) to the Laboratori Nazionali del Gran Sasso (LNGS) network. The DAQ server also interfaces to the DAQ components, including hardware configuration of the WFDs, GDM, and CDMs, and communicates with the DAQ Slow Control Processors (SCPs, \(11\)). Slow-control devices, including Cabinet Distribution Units (CDUs), VME crates, and Photon Detection Unit (PDU) power supplies, are managed through the control network switch (\(12\)).}
                \label{fig:daq-network}
            \end{minipage}
    \end{minipage}
\end{figure*}

\autoref{fig:daq-network} presents a schematic overview of the \gls{ds20k} \gls{daq} system and the network that interconnects its components. The lower part of the diagram shows the hardware elements installed on top of the cryostat, while the upper part presents the elements located in the IT room near the detector.
Starting from the bottom of the diagram, the custom-built Global Data Manager (\gls{gdm}, 1) and Crate Data Manager (\glspl{cdm}, 2) boards are shown. These boards are interconnected and distribute clock and command signals to the \glspl{wfd} serving the veto systems (blue and green, 3) and the \gls{tpc} channels (red, 4) (see \autoref{sec:gdm_cdm}).

Each digitiser is connected through a network switch (\(5\)) to dedicated \gls{fep} computers (\(6\)), where individual digitised waveforms are processed in software to identify peaks, or \textit{hits}, and extract relevant signal information (see ~\autoref{sec:fep}). 

\glspl{wfd} are connected to the switch via 10 Gigabit Ethernet (10 GbE) Direct Attach Cables, while optical Multi-mode cables connect the \gls{wfd} network switches to the \glspl{fep}, which are in turn connected to the main data switches (\(7\)) via \SI{10} GbE Direct Attach Cables. 

The maximum 
data output rate of each \gls{wfd} is dictated by the speed of its network interface, which can use \SI{10} Gbps standard, far exceeding the experiment's needs. The system has been tested to run sustainably at \SI{250} MB/s per digitiser, with the main limitation being the \glspl{fep} data processing (see~\autoref{sec:quadrant}).

To optimise data 
flow in the \gls{ds20k} \gls{daq} system, \glspl{wfd} employ onboard digital filtering and a time-over-threshold algorithm to identify 
interesting waveform segments. Only waveform segments containing at least one photoelectron are transmitted to the \glspl{fep}. This process, managed by the FPGA-based Dynamic Acquisition Window algorithm, efficiently discards waveform samples lacking relevant physics information, significantly reducing data volume without compromising signal integrity (see~\autoref{sec:digit}). Monte Carlo simulations indicate that the aggregate expected data rate from a single digitiser for the TPC has a strong dependence on the actual dark count rate of the photosensors. Assuming a dark rate of \SI{400}{Hz} per channel, a factor 10 in excess of what has been measured in vacuum with pre-production SiPMs, the simulation predicts about \SI{60} MB/s per board, comfortably within the rate mentioned above. Additional compression techniques and/or waveform downsampling in firmware can provide further data reduction if needed.

Next the data are transferred to the \glspl{fep}, where waveforms are processed and only the time, charge and prominence of the hits identified within the waveform are retained. This approach minimises data volume while preserving necessary event information for offline analyses. Full waveforms can be additionally saved for debugging purposes if needed.
The detector layout imposes the signal collection to be in four distinct locations or \textit{quadrants}, one for each chimney (see~\autoref{sec:daq_infrastructure}). Within a quadrant \glspl{pdu} can be assigned to any digitiser. Special mappings are left to further optimisation.

From the data switches, data are distributed via Direct Attach Cables to a cluster of Time Slice Processors (\glspl{tsp}, \(9\)), where the entire detector data from a predefined time period (Time Slices, see \autoref{sec:time_slice_architecture}) are merged and where online reconstruction is performed, enabling further data reduction.

Each \gls{tsp} processes all detector channels for a given \gls{ts}. The \glspl{ts} produced by the \glspl{tsp} are then collected by the Merger machine (10), which assembles them into a time-ordered sequence and stores the resulting data stream on disk (see \autoref{sec:tsp_and_merger}). The data stream can be delivered to an additional, dedicated machine that performs online processing of pre-selected data fragments for a combined physics analysis across all subdetectors, possibly contributing to the Supernova Early Warning System (SNEWS 2.0)~\cite{Al_Kharusi_2021}. 

The \gls{daq} server (\(8\))  has the critical task of orchestrating the data traffic between the digitisers and the \glspl{tsp}, the timing distribution and the acquisition sequences. This functionality is implemented by running \gls{midas}, a publicly available, general-purpose software used in several small- and medium-scale physics experiments ~\cite{842578,10126097,5750358}, through the \gls{midas} Supervisor. This \gls{ds20k} \gls{midas} server's role is to interface to all the \gls{daq} components, integrate communication, allow hardware configuration (\glspl{wfd}, \gls{gdm}, \glspl{cdm}) and communicate with the \gls{daq} \glspl{scp} (\(11\)). This includes the management of \glspl{cdu} (11), VME crates, and \gls{pdu} \glspl{ps} through a control switch (\controlswitch, \(12\)).

Finally, the data are transferred for storage through the Gateway machine (\(13\)) connected to the external \gls{lngs} network.

\subsection{DAQ infrastructure}
\label{sec:daq_infrastructure}
The \gls{gdm} and \gls{cdm} boards, along with the digitisers and the network switch are located on top of the detector cryostat and grouped into \(4\) racks placed close to the chimneys equipped with signal feedthroughs.
This location provides the shortest possible length of the signal cables to minimise noise pickup and signal integrity issues. Some elements of the \gls{dcs} and specific safety interfaces will also reside in this area.
All remaining machines are located in the \gls{ds20k} counting room, away from the detector.

The racks on the detector rooftop are standard closed 42U, \(600\)$\times$\(1000\) racks. Their \SI{1000}{\milli\meter} depth provides cabling space for the equipment.
Dedicated fans will provide ventilation with ambient forced filtered air. The inner rack airflow temperature can be monitored with external temperature sensors available in the VME crate as well as at each of the \glspl{wfd} (board and FPGA).

The \gls{daq} racks hosting the main \gls{midas} servers, network switches, \glspl{fep}, \glspl{tsp}, and \gls{dcs} equipment will be placed in the IT room on the side of the detector infrastructure.

\section{Time Slice Concept}
\label{sec:time_slice_architecture}

The \gls{daq} system of the \gls{ds20k} experiment is designed to operate in a fully triggerless mode, wherein each channel functions independently, continuously generating data without relying on a global trigger to initiate the acquisition. Instead, a local time-over-threshold logic at the single channel level is used to identify all waveform segments containing signal, without interrupting data acquisition.

\begin{figure*}[ht!]
    \centering
    \includegraphics[width=\linewidth]{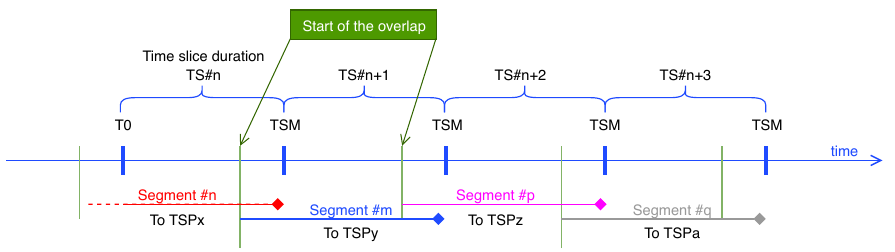}
    \caption{\gls{ts} concept: the acquisition timeline is segmented into \timeslice\ intervals, each directed to a dedicated \gls{tsp} for further analysis. Since \glspl{tsp} lack access to previously processed slices, a portion of each \gls{ts} (\drifttime) is duplicated and sent to the next \gls{tsp}. Time Slice Markers (TSMs) ensure proper segment assembly (see \autoref{sec:gdm_cdm}).}
    \label{fig:time-slice-concept}
\end{figure*}

Data selection for permanent storage occurs after signals from all detector components have been gathered into a single location and time sorted into data blocks called \glspl{ts}. ~\autoref{fig:time-slice-concept} shows a pictorial representation of the \gls{ts} timing. In this approach, the acquisition timeline is partitioned into intervals, each assigned later to a dedicated \gls{tsp} for further analysis. Upon completing the processing of a \gls{ts}, the \gls{tsp} signals its readiness to handle the next one.

Due to the analysis time, a single \gls{tsp} processes non-consecutive \glspl{ts}. As a consequence, it cannot handle physics events that span between two neighbouring \glspl{ts}. Such events require special handling. To address this issue, the end portion of each \gls{ts} is duplicated and forwarded to the next \gls{tsp}, ensuring boundary events are properly captured. The overlap corresponds to the maximum electron drift time in the \gls{tpc}, approximately \drifttime. Given a \gls{ts} duration of \timeslice, this overlap results in around 0.5\% of the analysed events being duplicated at the DAQ output stage. 
A \timeslice\ \gls{ts} is selected to detect supernova events within at most a few \glspl{ts}. Choosing a significantly smaller \gls{ts} would result in a larger volume of duplicated data to process. However, the \gls{daq} architecture makes it straightforward to adjust this parameter.

\section{Waveform Digitisation and Filtering}
\label{sec:digit}

Waveforms from the entire detector are transmitted as a differential signal to \(48\) commercial \digmodel\ CAEN \SI{16}{\bit}, \SI{125} MS/s, high channel density (\(64\) channels), \SI{4}{\volt} peak-to-peak, \SI{20}{\mega \hertz} bandwidth waveform digitisers with \(16\) programmable analogue gains~\cite{wfd5}. The \glspl{wfd} are placed in \(4\) crates, each containing \(12\) modules. Within a single crate, \(9\) out of \(12\) digitisers collect data from the \gls{tpc} while the remaining \(3\) are used for the inner and outer veto.

The digitisers support the integration of custom firmware, which can be uploaded to a reserved section of the FPGA using the OpenFPGA service~\cite{openfpga}. This feature grants direct access to the raw digital data stream, allowing for the implementation of custom acquisition control flows and data processing tasks such as triggering, data filtering and compression. Once these tasks are executed, control is returned to the CAEN firmware for data transmission via Ethernet to the \glspl{fep}. 

The core FPGA in the \digmodel\ is a Xilinx ZU19EG~\cite{fpga_model}. This device has a Quad-coreARM\textsuperscript{\textregistered} Cortex\textsuperscript{\texttrademark}-A53 MPCore\textsuperscript{\texttrademark} up to \SI{1.5}{\giga \hertz}. The ARM System-on-chip manages the interface from the Programmable Logic to the Programmable System running Petalinux.

The digitiser data output format is 64-bit aligned, and the first five words are reserved for the header. The header includes multiple fields accessible to the user. The digitised waveform samples follow the header, each consisting of \SI{16}{\bit} samples stored sequentially. 

The primary goal of the custom firmware is to implement a tailored trigger algorithm for identifying segments of the digitised waveforms containing at least one pulse corresponding to one or more photoelectrons. This process begins with the raw waveforms being processed through a \(64\)-coefficient, \(16\)-bit per coefficient \gls{fir} filter, used to suppress high-frequency noise and maximise the signal-to-noise ratio. Each channel uses \(16\) \glspl{dsp} to construct the \gls{fir} filter. 

Due to the limited available number of \glspl{dsp}, and since some of them are needed for other functions, the input signal is processed at an effective sampling rate of \SI{62.5} MS/s while the filter's \glspl{dsp} run at \SI{250}{\mega \hertz}, to reduce the filter's \gls{dsp} usage.
This configuration provides \(4\) filter clock cycles per input sample, enabling the application of 4 different coefficients per input sample over each sample interval. This allows the equivalent number of coefficients for each channel's filter to be \(64\), despite only \(16\) \glspl{dsp} being used per filter. Each of the filter's output samples is then extended over \(2\) clocks, effectively reverting the original \SI{125} MS/s signal rate.

Segments are retained if the signal exceeds the requirement for the duration of the amplitude above the threshold. 
Acquisition continues until the signal drops below a secondary, independently configurable threshold. A small post-trigger region is appended to each segment to extract key quantities, such as noise levels, while a short pre-trigger region captures baseline information. Finally, these complete signal segments are transferred to the \glspl{fep}~\cite{zlefw}.

This gated acquisition method maximises digitiser throughput while maintaining high peak detection efficiency. If a subsequent signal triggers the threshold during an ongoing acquisition, the gate is extended to include the additional signal and its post-trigger region. The gate extension is capped at a few tens of microseconds to avoid excessive data accumulation. Upon reaching this limit, the firmware truncates the waveform and resumes acquisition for subsequent segments once the signal falls again below the threshold. 

Waveform segments exceeding the minimal duration threshold are divided into smaller sub-segments to accelerate data transfer due to time sorting in the factory-loaded firmware implementation. These sub-segments are reassembled at the \glspl{fep} stage (\autoref{sec:fep}).

All the parameters needed for the configuration of the boards can be modified by the user through the \gls{midas} webpage. A non-exhaustive list of the main configuration parameters, their size and their function is provided in~\autoref{tab:wfd-configuration}.

\begin{table}[ht!]
    \centering
    \begin{tabular}{p{0.30\linewidth}p{0.27\linewidth}p{0.35\linewidth}}
    \toprule
         \textbf{Parameter}& \textbf{Size / range}&\textbf{Function} \\
         \toprule
         Readout channel mask & two 32-bit words & Select active channels\\
         Pre-trigger & 12-bit, per channel, in samples & Number of samples saved before the trigger condition\\
         Post-trigger & 12-bit, per channel, in samples & Number of samples saved after the signal falls below the post-trigger threshold\\
         Max segment length & 16-bit, in 4-sample units & Maximum segment length before splitting\\
         Load pattern & 1-bit flag & Load a collection of waveforms into the board\\
         Enable decimation & channel bit mask & Enable waveform decimation\\
         Decimation factor & integer word & Decimation factor\\
         Trigger threshold & 16-bit, per channel, ADC counts & Primary trigger threshold\\
         Post-trigger threshold & 16-bit, per channel, ADC counts & Threshold used to close the dynamic acquisition window\\
         Time over threshold & 16-bit, per channel, in samples & Minimum time above threshold required to accept a trigger\\
         Enable FIR filter & two 32-bit words & Enable onboard FIR filter\\
         FIR coefficients & 64 signed 16-bit words & FIR filter coefficients\\
         \gls{cdm} veto enable & channel bit mask & Enable veto from \gls{cdm}\\
         WAVE FIFO almost full & 16-bit, in FIFO words & Busy threshold for the waveform FIFO\\
         PARAMS FIFO almost full & 16-bit, in FIFO words & Busy threshold for the parameter FIFO\\
         DC offset & per-channel value & DC offset for the 64 channels\\
         \bottomrule
    \end{tabular}
    \caption{Main configuration parameters for the CAEN \digmodel\ modules. The sample-based parameters are expressed in digitiser samples.}
    \label{tab:wfd-configuration}
\end{table}

\begin{figure*}[ht!]
    \centering
    \includegraphics[width=\linewidth]{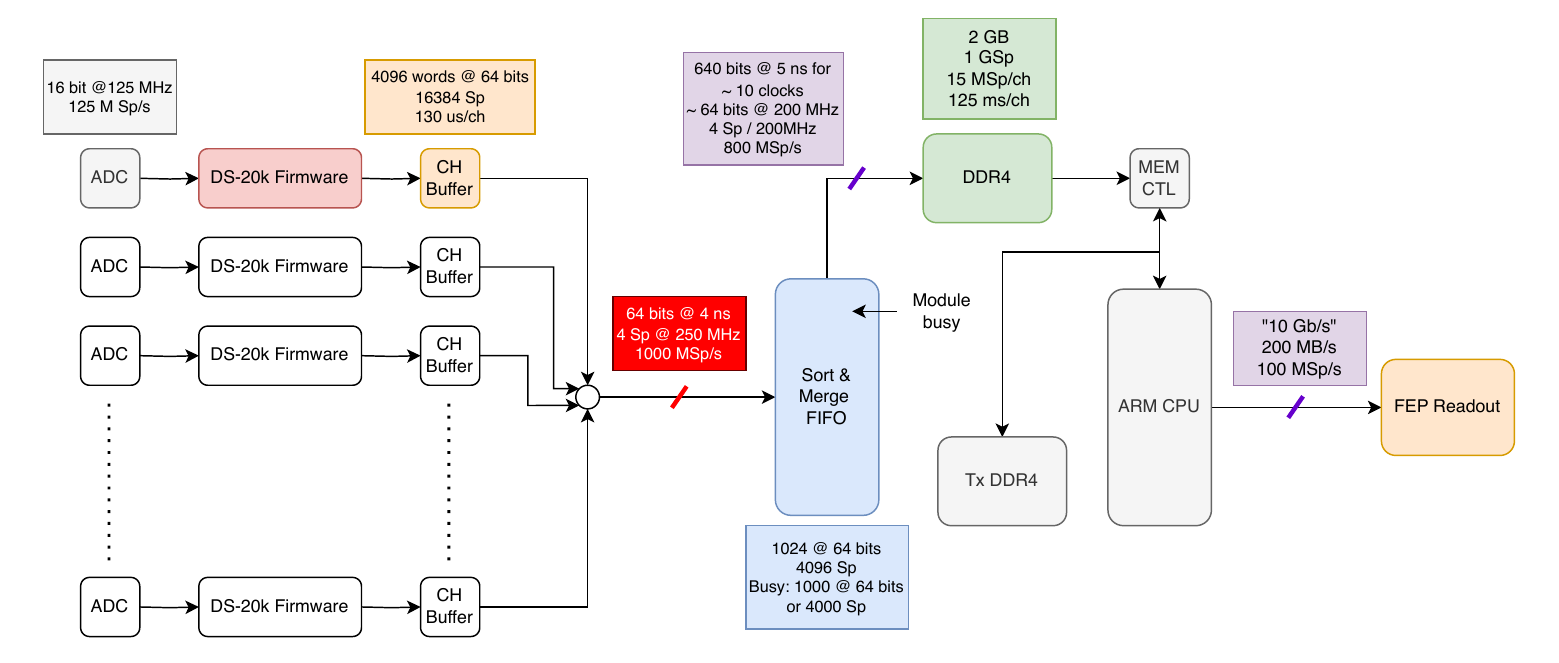}
    \caption{Schematic representation of the data path within the \digmodel\ boards. 
    The Sort \& Merge buffer collects data from all the 64 channels (CH) with a speed of \SI{64}{bit} at \SI{4}{\nano \second}, \(\SI{250}{MHz}\). Since each 64-bit word contains four 16-bit ADC samples, this corresponds to an effective throughput of 1000 MS/s, i.e. eight times the ADC sampling rate of 125 MS/s. Waveform segments are then transferred to a \SI{2}{\giga\byte} DDR4 memory with a speed of \SI{640}{b} over about 10 clocks of \SI{200}{\mega \hertz} then to an ARM CPU and finally to the \glspl{fep}.  }
    \label{fig:busy-scheme}
\end{figure*}

Each channel of the \digmodel\ features two \gls{fifo} memory buffers: the PARAMS buffer and the WAVE buffer, the latter shown along with the subsequent stages of the data path inside the digitiser board in~\autoref{fig:busy-scheme}. The PARAMS buffer stores metadata, including the channel number, timestamp, waveform size (in samples) and user-defined parameters. With a maximum depth of \(512\) 64-bit words, it handles single-variable entries and is highly unlikely to overflow.

The WAVE buffer holds the waveform data with a maximum capacity of \(4096\) words (\(16384\) samples). This buffer is susceptible to filling up during highly energetic or burst-like events, potentially creating a bottleneck for the system's data acquisition.

In addition to the per-channel PARAMS and WAVE, the VX2745 firmware implements a shared Sort \& Merge buffer in the FPGA logic. This buffer collects data from all the 64 channels with a speed of \SI{64}{bit} (4 samples) per \SI{4}{\nano \second}, (\SI{250}{\mega \hertz}). Since each 64-bit word contains four 16-bit samples, the corresponding effective throughput is 1000 MS/s, i.e. eight times the ADC (Analog-to-Digital Converter) sampling rate of 125 MS/s of the VX2745 modules. Measurements show, that in practice, due to limitations in the current firmware the achieved throughput is only a factor of five higher than the ADC readout rate, which may be lifted in future revisions.
Waveform segments are then transferred to a \SI{2}{\giga\byte} DDR4 memory with a speed of \SI{64}{bit} per \SI{5}{\nano \second} (\SI{64}{bit} at \SI{200}{\mega \hertz}).

\begin{figure}[ht!]
    \centering
    \includegraphics[width=0.7\linewidth]{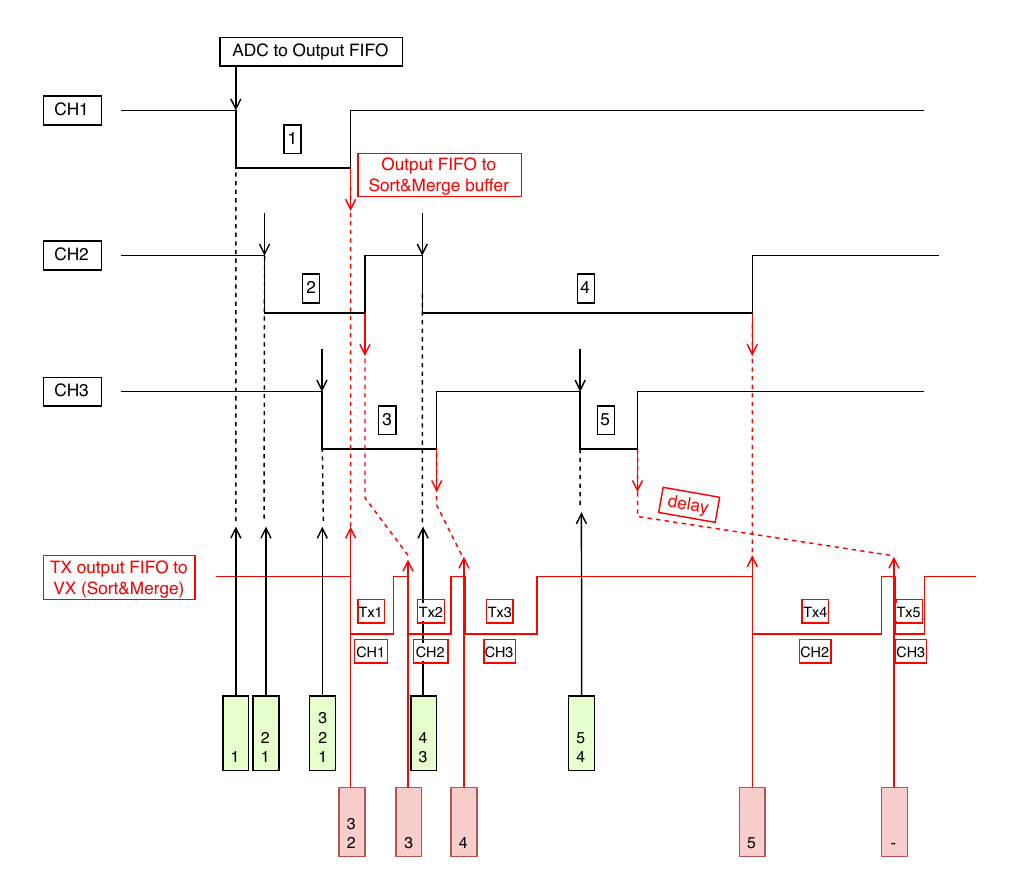}
    \caption{Schematic representation of the busy logic within a digitiser module. Black lines represent the data readout, while the red line represents the data transmission. Segment start and end times are identified by black and red arrows, respectively.}
    \label{fig:busy}
\end{figure}

A waveform segment is read as soon as its start time has been detected, and can only be transferred after its end time has been recorded. 
If a waveform remains above the trigger threshold, subsequent segments with start times later than the ongoing waveform’s start time are queued in a separate \gls{fifo}, regardless of their end times. For this reason, long waveform segments can delay the transfer of all earlier segments that have already fallen below the threshold but started after the ongoing waveform. These delayed transfers can quickly fill the \gls{fifo} queue, posing a channel buffer overflow risk and requiring the management of the \gls{fifo}'s own ALMOST FULL signal at the system level to avoid data loss. ~\autoref{fig:busy} shows a schematic representation of the busy logic within a digitiser module. Black lines represent the data readout, while the red line represents the data transmission. Segment start and end times are identified by black and red arrows, respectively.
A busy condition is asserted by the OR of individual channel ALMOST FULL signals. Upon asserting a busy, the signal travels to the corresponding \gls{cdm} and back to the \gls{gdm}, which suspends the acquisition for all the \glspl{wfd} at the same time until the FIFOs are emptied. Finally, waveforms are transferred to the ARM CPU and the \glspl{fep}. 

\subsection{Simulation and validation}
\label{sec:busy_sim_validation}

There are several possible sources that could cause the digitiser board to assert a busy condition. At the module level, this might be caused by the backlog from the \glspl{fep} due to transmission or filtering. In this case, the readout is paused. If too large a portion of the DDR4 buffer is filled and the readout pause is long enough, then the Sort \& Merge Buffer may assert a module busy.

At the individual channel level, long pulses on some channels preventing later channels from transferring their content to the Sort \& Merge Buffer can cause an accumulation of pulses in the other channels' \glspl{fifo} with a subsequent overflow. This is considered the main source for asserting a module-level busy condition and has been studied with the help of a detailed simulation.

The software emulates a collection of \(64\) input waveforms generated by the \gls{ds20k} Monte Carlo (MC) background simulation and evaluates instantaneous WAVES buffer occupancy for each channel at a sampling period of \SI{8}{\nano \second}~\cite{g4ds}. A sample corresponding to \SI{10000}{\second} of full \gls{ds20k} background MC was processed. The \SI{1}{\second} \glspl{ts} were divided into \SI{1}{\milli\second} sub-slices to make the calculation tractable while preserving the time scale relevant for FIFO filling. For each sub-slice, groups of \(64\) neighbouring channels were selected to emulate the input of one digitiser module.

The simulation output was validated by cross-checking it against the same input fed to the digitiser using a special firmware version. This firmware implementation allows the loading of up to \SI{1}{\milli\second} of MC data in the digitiser at the nominal sampling rate \SI{125} MS/s. Additionally, it provides analogue signals monitoring individual channel buffer occupancies, allowing for direct confirmation and validation of the software busy emulation.

On average the simulation predicts a maximum channel FIFO occupancy around 20\%. The risk of filling the FIFO comes from burst-like topologies, such as large or overlapping S2 signals, that can provoke FIFO overflows. The busiest digitiser exceeds an \(80\%\) FIFO-occupancy threshold in about 4\%  of \SI{1}{\milli\second} sub-slices. When this occurs the almost-full FIFO is drained to an acceptable level in about \SI{1}{\milli\second}, thus producing a 4\% dead-time for the entire system.

Several mitigation strategies were studied using the same simulation framework. Splitting long waveform segments into shorter sub-segments expedites data transmission by allowing later, shorter segments to be transferred before the full long segment has completed. A split length of 628 samples was tested and found to reduce the maximum buffer occupancy by less than 20\% at the \(80\%\) FIFO threshold. Firmware compression has a larger impact: a factor-two reduction of the waveform payload gives an almost 50\% reduction in the simulated buffer occupancy.

The splitting and downsampling algorithms have been implemented in firmware and tested.
The implementation of compression is under development. A fast, two‐stage, lossless compression algorithm has been implemented in VHDL and integrated into the Open FPGA firmware. The compression factor based on the \gls{ds20k} \gls{pdu} waveform data has been measured to be better than 2. The algorithm combines first‐order differential coding (delta coding) with Huffman entropy coding~\cite{Huffman}. First, each sample is replaced by its residual with respect to the previous one. This reduces dynamic range as the most probable values are small, i.e. the probability concentrates near zero. In the second stage these residuals are encoded using Huffman code: each value in the range [-64, +64] is assigned a variable‐length bit pattern according to its frequency in the typical waveform. Less frequent, out‐of‐range values start with the “escape” code followed by a raw 16-bit value. This means that a \gls{lut} must be first created based on a sample of expected waveforms.
The \gls{lut} is indexed by the residual value. Each item contains two numbers: the length in bits of the encoded value and an upshifted bit template (padded with zeros to the required length in bits if needed).

At runtime, the encoder reads four data samples in two clock cycles, computes residuals, and accesses the \gls{lut} to retrieve codewords and their bit lengths. These are then concatenated and written to a small register buffer. When 64 bits (the equivalent of four samples of raw data) are accumulated in the buffer, they are transferred to the channel data \gls{fifo}. The buffer is then cleared and filled with overflow bits (if any). The coding operations are performed across four successive pipeline stages.
The decompression algorithm is written in \CC\ and has been tested on the \gls{fep} machines. It can decode compressed data at the rate of \SI{230} MB/s per core.
The development of this algorithm allows the \gls{daq} system to mitigate the potential busy issue, extend the maximal waveform segment length and increase the data rate from the digitiser to the \gls{fep}.

\section{Global Data Manager and Crate Data Manager}
\label{sec:gdm_cdm}

\begin{figure}[tpb]
    \centering
    \includegraphics[width=0.7\linewidth]{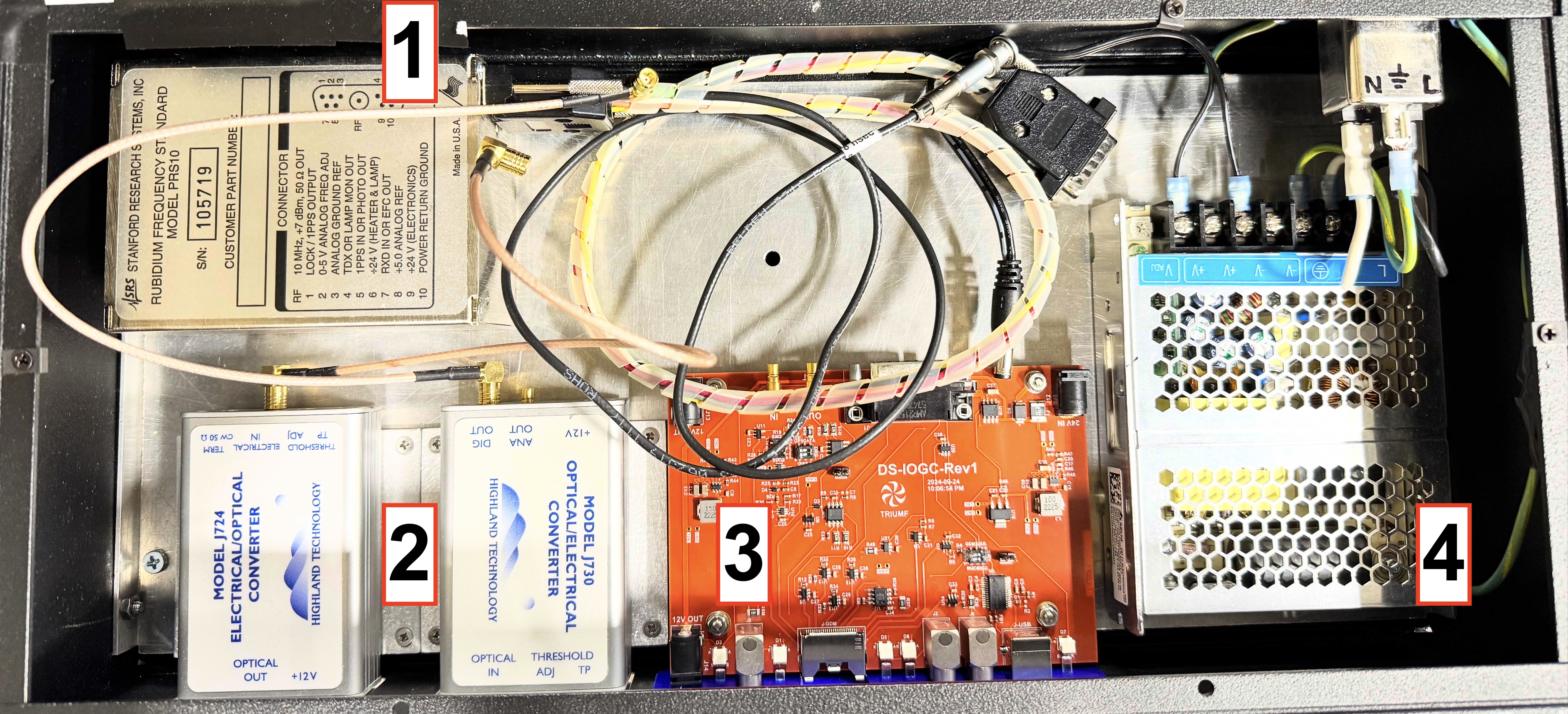}
    \caption{The picture shows a crate box with Stanford Research System PRS10 rubidium clock (\(1\)), optical/electrical and electrical/optical converters (\(2\)), a custom input-output GPS Clock board (IOGC, \(3\)) developed in TRIUMF and a power supply (\(4\)). The synchronisation signal is delivered by the \gls{lngs} laboratory by optical fibre; the signal is converted to electrical and delivered to the rubidium clock through the IOGC board, which in turn is connected through the Samtec connector to the \gls{gdm}. The electrical/optical converter will be used for calibration by sending a signal through \gls{lngs} infrastructure, allowing for precise measurement of the synchronisation signal delay. }
    \label{fig:clk_box}
\end{figure}

Because of the large channel number used in DS-20k multiple digitisers must work concurrently. A shared sampling clock is essential to ensure precise synchronisation in phase and time across all digitisers. Precise absolute time, while not relevant for a Dark Matter search with DS-20k, is required for correlation of supernova burst events, with signals detected by other detectors worldwide, while phase synchronisation of all channels is necessary for proper event reconstruction.

A synchronisation signal packet is delivered by the \gls{lngs} to the underground laboratory via optical fibre to allow for assigning accurate timestamps to events by all experiments~\cite{Deo_2019}. This packet is constructed at the surface facility and is based on a GPS signal. The packets are sent at 1 Hz rate and the first edge marks the synchronisation time. Each data packet comprises the absolute time of the previous packet and the GPS clock bias correction expressed in nanoseconds.
This signal is fed to a custom input-output GPS Clock board (IOGC, see ~\autoref{fig:clk_box}) and disciplines a local atomic clock (Stanford Research System PRS10 rubidium frequency standard~\cite{PRS10}). The system is able to provide absolute time with an accuracy of \SI{15}{\nano \second}. In case of loss of the \gls{lngs} synchronisation signal, the PRS10 clock can hold a Stratum 1 level for \(72\) hours, namely the next level in precision below GPS.

\begin{figure}[tpb]
    \centering
    \includegraphics[width=0.7\linewidth]{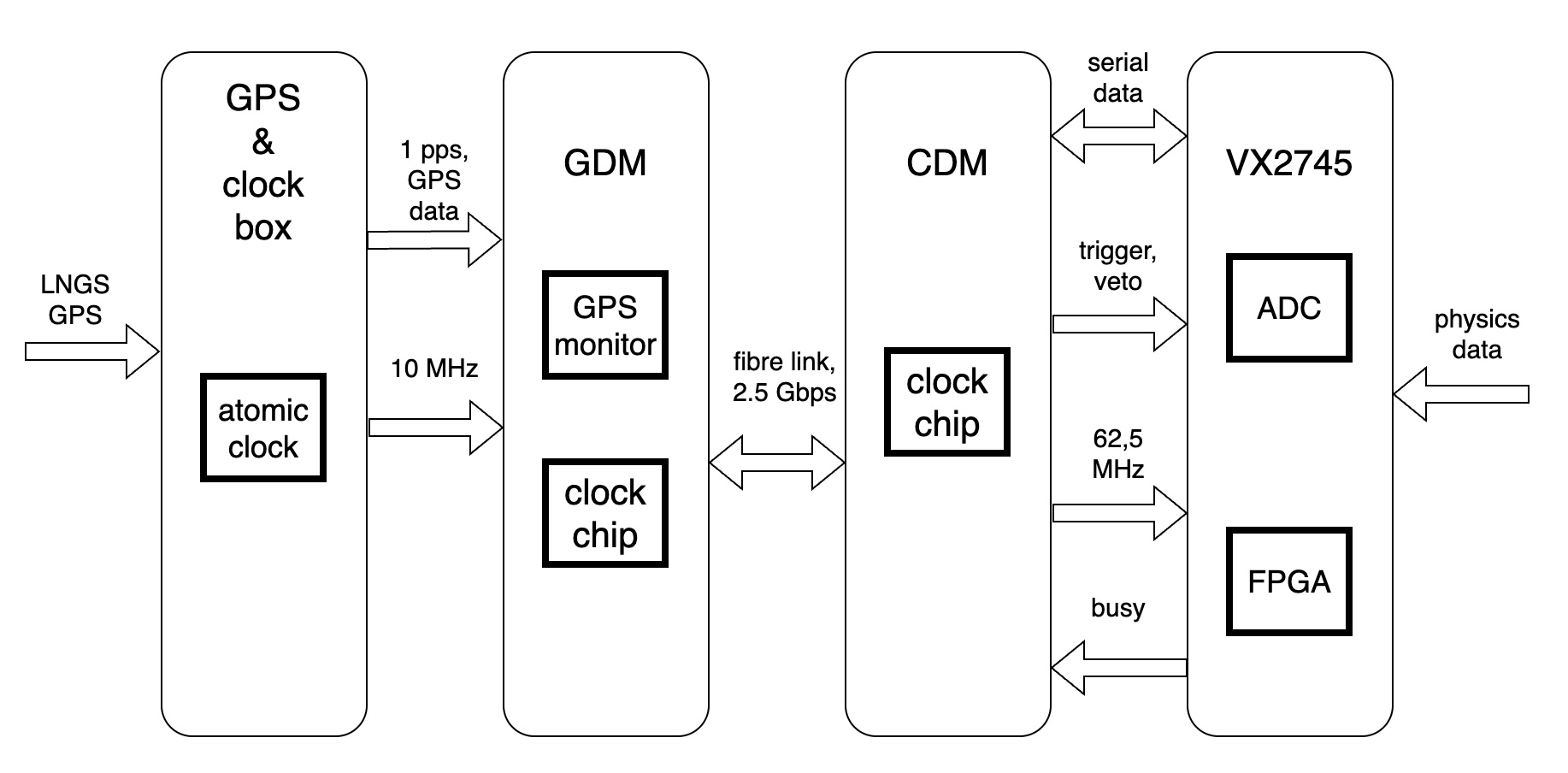}
    \caption{Schematic representation of the clock distribution and data management with the \gls{gdm} and \gls{cdm} boards. From the left: the timing information is delivered to the underground laboratories by \gls{lngs} with the optical fibre signal formed in the surface laboratory based on the GPS data. This signal is used to drive the local Rb clock. The \gls{lngs} stream is converted to electrical and, in parallel with 10 MHz signal from the atomic clock, transferred to the \gls{gdm} located nearby the clock box. The \gls{gdm} decodes the absolute time correcting for the signal delay from the surface laboratory. The optical links operating at \SI{2.5} Gbps are used to distribute the phase-aligned recovered clocks and time-synchronous control packets to 8 \gls{cdm} boards which in turn send it to all the 48 digitisers.}
        \label{fig:cdm-scheme}
\end{figure}

~\autoref{fig:cdm-scheme} displays a schematic representation of clock distribution and data management in \gls{ds20k}. Two custom data manager boards were developed at TRIUMF for \gls{ds20k}: the \gls{gdm} and \glspl{cdm} (1 and 2 in~\autoref{fig:daq-network} respectively). The reference \SI{10}{\mega \hertz} clock maintained by PRS10 atomic clock together with 1 Hz \gls{lngs} synchronisation signal are delivered to the \gls{gdm} board. The single \gls{gdm} board is connected to 8 \glspl{cdm}. In each crate, 1 \gls{cdm} controls digitisers collecting data from the \gls{tpc} (\(9\) out of \(12\)), and the second one controls the remaining three digitisers used for the inner and outer veto. 

This design results in the \gls{daq} infrastructure having a unique clock distributed from the \gls{gdm} to the digitisers through multiple \glspl{cdm} located near the different groups of digitisers. The communication between the \gls{gdm} and the \glspl{cdm} occurs via an optical link operating at \SI{2.5} Gbps. Optical links are configured in real time to distribute phase-aligned recovered clocks and time-synchronous control packets to all digitisers. The synchronisation between the digitisers is achieved by distributing a single, highly stable, phase-aligned clock signal to all data converters. The phase shift of the data between the channels within a single \gls{wfd} and between \glspl{wfd} has been measured by feeding an identical sine wave signal through a waveform generator connected to a fan-out. The maximum time shift is measured to be below \SI{500} ps for any channel and digitiser pair.

The synchronised time segmentation mechanism requires transmitting a \gls{tsm} to all the digitisers to ensure proper segment assembly. The data packets transmitted from the \glspl{cdm} to the digitisers are decoded and reformatted for communication at a frequency of \SI{125}{\mega \hertz}, which is dictated by the limitations of the digitiser hardware components. A phase-aligned \SI{62.5}{\mega \hertz} clock, derived from the \SI{125}{\mega \hertz} source, is routed to the front panel of the digitisers, serving as the main clock. 

Full-duplex communication is used to transmit various critical data packets. reference clock, \gls{tsm},  external trigger , reception of the real-time hit map from all channels, and operation control.  
The real-time hit map is a detector-wide, one-bit-per-channel record indicating whether the signal exceeded threshold during a \SI{1.2}{\micro\second} snapshot; it is made available to the \gls{gdm} for hit-rate monitoring and for generating hit-map-based triggers. The hit-map transmission time plus trigger latency must fit within the \SI{8}{\micro\second} input ring buffer. Operation control refers to commands sent by the \gls{gdm} to the \glspl{wfd}, including acquisition enable/disable and distribution of the 48-bit trigger word.
In particular, the \gls{tsm} is utilised during the merging of data fragments in the \glspl{fep}, specifications regarding the trigger type and sector address are essential in case of triggered operation of the \gls{daq} system; here, a sector corresponds either to one of the 48 digitisers or, equivalently, to one of the 48 bits composing the trigger generated by the \gls{gdm}. 
A single bit in the control packet is used as a veto to suspend and resume data acquisition to prevent data loss. For this purpose, each \gls{wfd} provides a busy signal that is asserted when its buffer occupancy reaches a predefined threshold (\autoref{sec:digit}). This mechanism allows the system to react promptly, suspending acquisition across all digitisers to ensure complete and accurate \gls{ts} data collection, which is essential for proper event reconstruction.

The \gls{tsm} bit, along with locally stored and time-corrected \gls{lngs} time information, triggers a \gls{tsm} event in the \gls{wfd}. Each digitiser generates a single \gls{tsm} event marked with a bit in the header and the time information encoded in the data section. The \gls{tsm} event is transferred to the \gls{fep} and does not contain any waveform data. This event is essential for merging data fragments across all the digitisers and detecting any missing fragments within the \glspl{fep}. \glspl{fep} build \glspl{ts} for further processing from data occurring between two consecutive \glspl{tsm}. When a \gls{tsm} is detected, any data outside of the given \gls{ts} is considered missing.

Furthermore, the control packet includes command fields for external trigger requests originating from the \gls{gdm}, which can initiate the acquisition of the currently buffered data in the digitisers. These external triggers may stem from inputs available on the \gls{gdm}, such as test pulses, calibration devices or algorithmic decisions based on the hit-map received from the digitisers. The triggered and triggerless operations are mutually exclusive. The timing precision is provided by the common phase-aligned clock distributed through the \gls{gdm}/\gls{cdm} chain, as described above. Synchronisation at the data-flow level is maintained through the \gls{tsm} and busy/veto information, as discussed in Sections~\ref{sec:time_slice_architecture} and~\ref{sec:digit}. 

\section{Front End Processors}
\label{sec:fep}

The \glspl{fep} are responsible for several tasks. They acquire waveform data from the CAEN \digmodel\ digitisers and perform single-channel data processing, which includes single waveform segment digital filtering and data reduction. Additionally, the \glspl{fep} time sort the event data across all the connected \digmodel\ units (each \gls{fep} reads data from 2 digitisers) in time slices based on the \gls{tsm} counter and listens for commands from the \gls{pm} application to initiate the transfer of the \gls{ts} data to the next idle \gls{tsp}.

These tasks run concurrently in separate threads to optimise CPU usage and network bandwidth. The communication between threads is based on software queues.
One thread per board is responsible for reading the data. Another thread processes the queue, subtracts baselines and applies filtering to all waveforms. Finally, an additional thread is responsible for building and shipping the \gls{ts}.

The \glspl{fep} are hosted on servers equipped with AMD 7700/7900-class CPUs and DDR5-4800 ECC memory, with \(2\times32\) GB DIMMs per machine. These machines were selected to provide high single-core performance and memory bandwidth for the main online tasks: receiving UDP data streams at rates up to about \SI{400}{MB/s}, decompressing waveform data when enabled, applying data reduction, time sorting the incoming fragments, and buffering several \glspl{ts} before transmission to the \glspl{tsp}. 

The primary physics function of the \glspl{fep} is waveform data processing, which involves extracting the time and prominence of all signal peaks (hits) within the segments received from the digitisers, where each peak corresponds to one or more photoelectrons.
The waveform data processing algorithm operates in steps to efficiently identify hits. The first one involves removing the baseline from the inverted raw waveform segment. The baseline values are calculated as the average of a predetermined fixed number of samples (150) in the first part of each waveform segment. Next, each waveform receives a calibration to equalise the gains across all the detector readout channels.
Then, the baseline-subtracted and calibrated waveform is processed using an infinite response Auto-Recursive exponential filter. 

The hit finding algorithm was developed to improve real-time hit identification and is based on the difference between the matched exponential filter and a moving average filter. The matched filter maximises the signal-to-noise ratio and produces sharp, cusp-like peaks at the hit positions. Subtracting a moving average of the filtered waveform suppresses slow baseline variations and correlated low-frequency noise, allowing individual hits to be clearly separated. Three key parameters govern the hit-finding algorithm: the moving-average window length (60 samples), the minimum integrated filtered-waveform response, and the hit-prominence threshold. Candidate hits are first required to have an integral of the filtered waveform above threshold greater than 12 samples, and then a prominence larger than 0.6 in single photoelectron units. These cuts are tuned to keep high single-photoelectron efficiency while reducing fake hits from noise fluctuations. A preliminary optimisation of the algorithm on simulated data indicates that 99\% efficiency on single photo-electron signals can be reached with O(10 Hz) of noise-induced fake signals.
The \glspl{fep} calculate the prominence, charge and time position of each hit. These, together with information on the individual waveforms (time, duration, integrated charge and number of identified hits) constitute the only information transmitted in standard data taking conditions to the subsequent \gls{daq} stages while the individual waveforms are discarded.

\begin{figure*}[ht!]
    \centering
    \includegraphics[width=\linewidth]{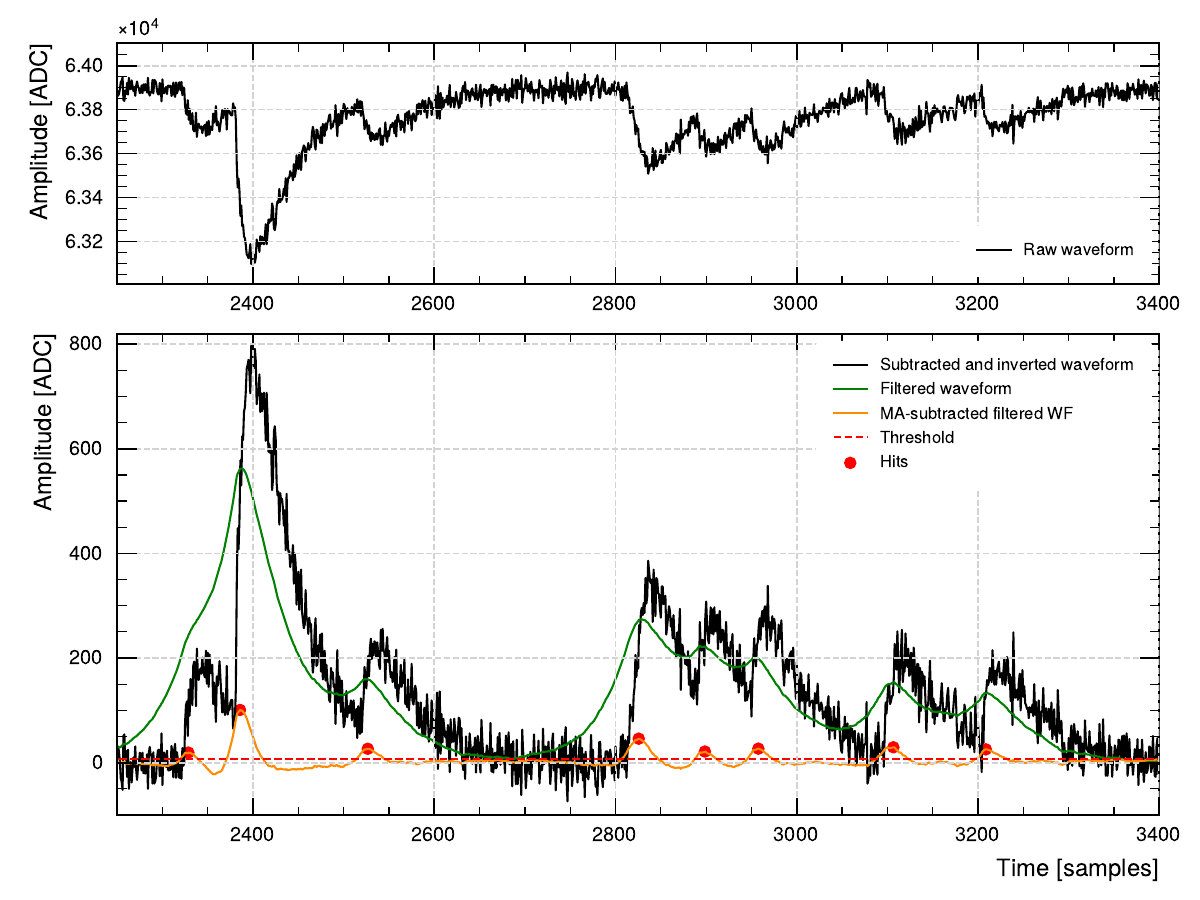}
    \caption{Waveform processing in the \gls{fep}. Baseline-subtracted waveforms are first inverted, then filtered using an Auto-Recursive exponential filter (green line), their moving average is subtracted (orange line) and then a peak finder is applied (red points). The peak finder looks for signals exceeding a preselected threshold in the moving average-subtracted waveform. }
    \label{fig:fep-processing}
\end{figure*}

~\autoref{fig:fep-processing} shows the waveform processing performed by the \glspl{fep} from the filtering to the hit finding stage.

Additionally, all the waveform segments over the full \gls{ts} duration coming from the \(128\) channels (64 per module) are decimated and separately summed for the channels belonging to the top and bottom Optical Planes with reference to the beginning of the \gls{ts} and encoded in \gls{zle} format~\cite{zle,zlealgo,dppcaen} to be transferred to the \glspl{tsp}, together with the information on hits and the summary information about individual waveform segments discussed above. This format enables a zero suppression of the regions where no signal has been detected, thus reducing the data throughput. In this respect, a further cut on the summed waveform amplitude might be applied at this stage.

Finally, the \glspl{fep} are responsible for monitoring the channels and assessing the raw data quality. A waveform and a reduced data stream are periodically extracted at the \gls{fep} level and transmitted to a remote analysis processor via the \gls{midas} event transfer system. 
This channel monitoring ensures data integrity while minimising the potential impact on extracting physics information. Various relevant metrics, such as prominence histograms, the integrated charge of raw waveforms and baselines, are continuously assessed and displayed on a dedicated \gls{midas} webpage during this monitoring process.

\section{Data Flow Control, Pool Manager and MIDAS supervisor}
\label{sec:pm}

\begin{figure*}[ht!]
    \centering
    \includegraphics[width=\textwidth]{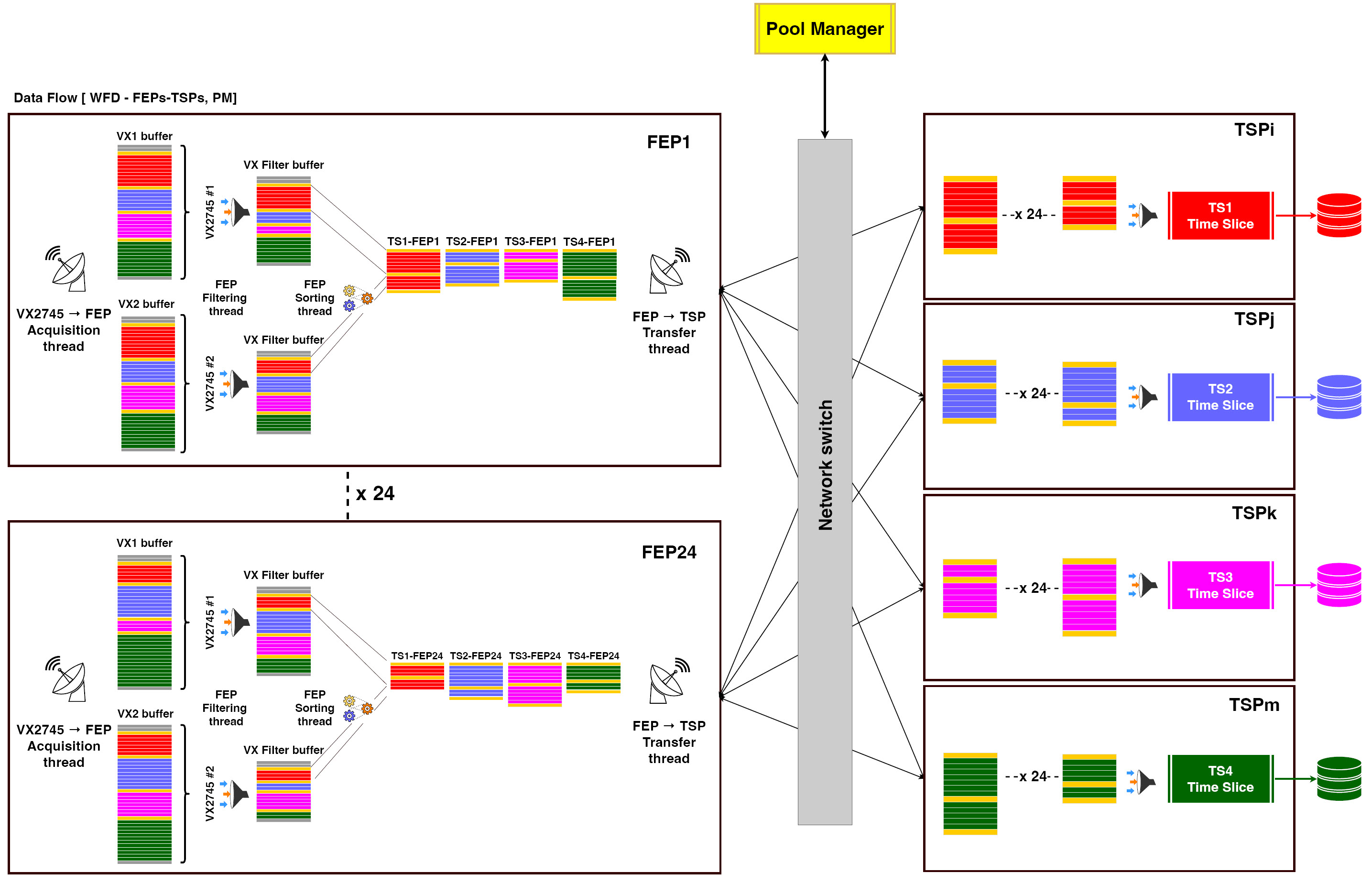}    \caption{Schematic representation of the data flow in \gls{ds20k}. Data are transferred from the waveform digitisers (\glspl{wfd}) to the Front End Processors (\glspl{fep}), where they are sorted in time and organised in \timeslice\ Time Slices (\glspl{ts}). In the picture, fragments having the same colour belong to the same \gls{ts}. \glspl{ts} are then sent from the \glspl{fep} to the \gls{tsp} through the network switch, under the Pool Manager's (\gls{pm}) supervision. The \gls{pm} assigns the appropriate \gls{tsp} address to each \gls{fep}, enabling the transmission of \gls{ts} data.}
    \label{fig:ds20k-data-flow}
\end{figure*}

Data transfer from \glspl{fep} to \glspl{tsp} is orchestrated by the Pool Manager (PM) application, which runs on the \gls{midas} server. The primary role of the \gls{pm} is to assign the appropriate \gls{tsp} address to each \gls{fep} for a given \gls{ts} transmission. A pictorial representation of the \gls{ds20k} data flow is displayed in~\autoref{fig:ds20k-data-flow}.

The \gls{pm} maintains an idle processor queue implemented as a \gls{fifo} data structure, which stores identifiers of the \glspl{tsp} (\gls{tsp}-ID) ready to process a new Time Slice. The \gls{pm} continuously scans this queue, retrieves the oldest \gls{tsp}-ID, and pairs it with the next available Time Slice-ID, forwarding this information to all the \glspl{fep}. Each \gls{fep} then stores the \gls{tsp}-ID and corresponding Time Slice-ID in its queue, allowing multiple \gls{ts} data segments to be simultaneously transferred to available \gls{tsp} destinations. This parallel processing scheme significantly optimises network bandwidth utilisation. 

Once data have been transferred to the \glspl{tsp}, the \gls{pm}, while waiting for the reception of the summary notification packets from any active \gls{tsp}, will direct the next \gls{ts} data packet to the next idle \gls{tsp}. 

Once a \gls{tsp} has completed the analysis of its assigned \gls{ts}, it is responsible for notifying the \gls{pm}. This communication includes a notification of the analysis completion status and a detailed report on the outcome of the most recent \gls{tsm} analysis.
The notification message is composed of a \gls{tsp}-ID (\gls{tsp} node ID), the \gls{ts}-ID, representing the number or identifier of the completed \gls{ts}, the transfer time, indicating the duration required to transfer the \gls{ts} from the \glspl{fep} to the \gls{tsp}, the size of the \gls{ts}, the time taken to analyse the slice, and the output data size, generated after the completion of the analysis.

Upon receiving the notification packet, the \gls{pm} processes it in two steps. First, the \gls{pm} adds the \gls{tsp}-ID to the idle processor queue, making the \gls{tsp} available for handling the next \gls{ts}. Second, the \gls{pm} composes an event status for the recently completed \gls{ts}. This status and related information are integrated into the \gls{midas} software infrastructure. They can be accessed via a custom web history page, allowing real-time monitoring and review of processing events.

Communication between the \gls{tsp}, \gls{pm}, and \glspl{fep} occurs through the central data switch using ZeroMQ (ZMQ) asynchronous messaging library, which provides an efficient method for message broadcasting~\cite{zmq}. Despite sharing the main data network, the communication overhead introduced by ZMQ is minimal and has a negligible impact on overall data transfer performance.

The \gls{midas} software package manages the overall data acquisition system. It performs several key functions: configuring the readout equipment, orchestrating the run sequence, generating various alarm levels (e.g., warnings, errors, or custom alarms) based on user-defined criteria, controlling data transfer to analysis tools, recording data for permanent storage, and managing equipment operation and monitoring. Additionally, \gls{midas} provides an application framework for device interfaces.

In the \gls{ds20k} experiment architecture, while \gls{midas} continues to handle overall control and monitoring, its role has been specifically adapted to manage data flow control only, rather than data transfer. This architectural decision enables raw socket data transfer from the digitisers to the \glspl{tsp}, allowing for optimal data throughput in the system.

This monitoring role also provides the bookkeeping needed to determine the \gls{daq} live time. When a digitiser buffer reaches the busy threshold, the corresponding \gls{wfd} asserts a busy signal, which is propagated through the \gls{cdm} to the \gls{gdm} and results in a global pause of the acquisition. In this context, the pause inhibits new self-triggered waveform segments in all digitisers, while data already accepted before the pause remain in the buffers and are drained downstream. Signals arriving during the pause are therefore not acquired and are accounted for as \gls{daq} deadtime. The pause remains active until the buffers drain below the recovery threshold, after which the \gls{gdm} sends a resume command. Busy, pause and resume transitions are counted by the \glspl{cdm} and timestamped by the \gls{gdm}; missed-trigger counters in the following data header provide an additional diagnostic of inhibited triggers.

\section{Time Slice Processors and Merger}
\label{sec:tsp_and_merger}

\glspl{tsp} manage incoming connections from the \glspl{fep}, receiving their data payloads. Once a \gls{ts} is processed, the \gls{tsp} notifies the \gls{pm} and waits for the next \gls{ts}.

The primary role of \glspl{tsp} is to collect and select hits to ensure efficient data processing. Additionally, they complete the sum of all the top and bottom channels by adding up the partially summed waveforms received from the \glspl{fep}.
The \glspl{tsp}' functionality can be extended to perform more advanced analyses, including event classification, and anomaly detection. 

Events can be categorized into different types, such as:
Regular DM (e.g. S1 in the WIMP region of interest), high energy gammas in the \gls{tpc} for calibration (e.g. high energy S1), low energy S2 events (e.g. low multiplicity S2 pulses isolated from other pulses), and Inner/Outer Veto (e.g., pulse above the threshold necessary for cosmogenic suppression and calibration or monitoring).  
This classification process allows the system to pre-scale specific event types, 
thereby reducing data storage requirements, if needed. By performing these analyses, the \gls{tsp} generates a new set of data that is stored locally on each \gls{tsp}'s storage device.
Time slices showing statistically significant anomalies, like an excess of low energy signals characteristic of a neutrino burst from a core-collapse supernova within the galaxy~\cite{DarkSide20k:2020ymr}, can be tagged for further quasi-online analysis downstream.

Each \gls{tsp} maintains its file composed of non-consecutive \glspl{ts} for the run period, while the \gls{pm} records the \glspl{ts} IDs processed by each \gls{tsp}. 
\glspl{tsp} notify the \gls{pm} at key stages of their operations: when the analysis of a \gls{ts} begins, if the analysis fails, and when the transmission of the \gls{ts} to the following data acquisition stage is completed. The \gls{pm} then forwards this information to the Merger, the machine responsible for collecting the \glspl{ts}. 
The specifications for the \gls{tsp} machines are similar to the \glspl{fep} (see~\autoref{sec:fep}).

The Merger receives the \glspl{ts} from the \glspl{tsp} via raw TCP/IP sockets,
sorts the \glspl{ts} chronologically and concatenates them into a continuous data stream. Once sorted, the \glspl{ts} are stored locally for temporary storage. In case of transmission failure between a \gls{tsp} and the Merger, the local \gls{tsp} copy is to be used to retrieve the \gls{ts}. In addition, the Merger handles the transfer of the concatenated \glspl{ts} to the \gls{cnaf} data centre for long-term storage and offline analysis. A \SI{100}{\tera\byte} local disk, corresponding to approximately one week of data taking, will be kept and deleted once the transfer has been acknowledged.

The Merger continuously monitors for missing \glspl{ts} during data collection. If a missing \gls{ts} is detected, it checks the \gls{pm} logs to determine the status of the corresponding \gls{tsp}. If the analysis has failed, the slice is marked as irrecoverable and no longer considered missing. However, if the analysis has started but not yet completed,  the Merger waits for a predefined time before proceeding to the next \gls{ts}.
\glspl{ts} not immediately delivered to the Merger remain temporarily stored on the corresponding \gls{tsp} until successful transmission, ensuring robustness against acquisition delays or failures without data loss.

Beyond sorting and concatenating \glspl{ts}, the Merger also delivers anomaly-tagged slices together with a predetermined number of preceding and following \glspl{ts} to a dedicated processor, possibly equipped with GPUs, implementing, e.g. a fast supernova trigger. An implementation prototype is currently under development. It will analyse, in quasi real-time, the collected \glspl{ts} for anomalous event rates across the TPC and Veto volumes. If a potential supernova event is detected, this processor can trigger an alert, that might be sent to e.g. to SNEWS~\cite{Al_Kharusi_2021}.

Since the Merger does not perform event reconstruction, but only file concatenation and bookkeeping, it can keep the output file open while querying the \gls{pm} and waiting up to a configurable timeout for delayed slices; the chronological ordering is then used offline to remove duplicated activity in the overlap between consecutive \glspl{ts} and to provide contiguous slices for supernova-burst follow-up.

\section{Slow Control}
\label{lab:slow-control}

\begin{figure*}
    \centering
    \includegraphics[width=\linewidth]{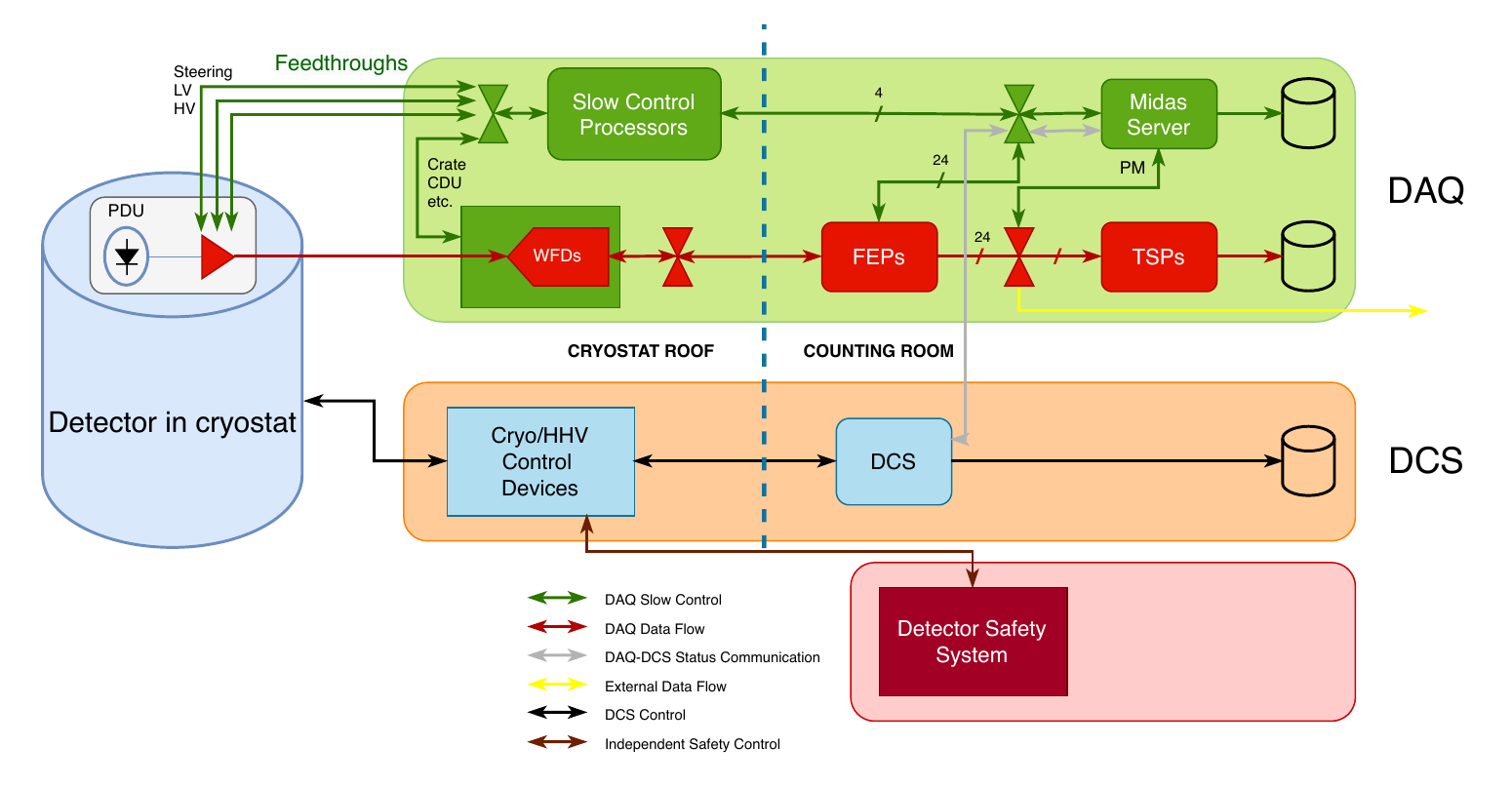}
    \caption{\gls{ds20k} Slow Control (\gls{sc}) system, structured in two complementary components: the \gls{daq} \gls{sc} and the \gls{dcs}. The \gls{sc} manages devices integral to the data acquisition process, including \gls{sipm} readout electronics, \gls{lv} and \gls{hv} supplies, steering control, and VME crates. The \gls{dcs} independently controls and monitors critical detector infrastructure, including the cryogenic system (both AAr and UAr) and \gls{tpc} high-voltage system for drift and electroluminescence fields. The blue dashed line separates the equipment that will be positioned on top of the cryostat from that in the experiment counting room.}
    \label{fig:slow-control}
\end{figure*}

The \gls{ds20k} Slow Control (\gls{sc}) system comprises two complementary components: the \gls{daq} \gls{sc} and the \gls{dcs}, as shown in~\autoref{fig:slow-control}. 

The \gls{sc} manages devices integral to the data acquisition process, including \gls{sipm} readout electronics, power supplies, and steering control. It is also used to oversee the \gls{daq} racks together with all its components. The \gls{sc} ensures real-time monitoring and regulation of the power and temperature of the digitisers, \glspl{cdm}, \gls{hv} and \gls{hv} power supplies, \glspl{cdu} and network switches. The \gls{sc} is operated by a dedicated computer and provides the network interface to the \gls{dcs}.

The \gls{dcs} independently controls and monitors the critical detector infrastructure, including the cryogenic system (both AAr and UAr) and the \gls{tpc} high-voltage system for drift and electroluminescence fields. 

Commercial \gls{hv} units provide the precision current monitoring required for individual channel I–V curve measurements, together with a ``steering" system for channel and \gls{pdu} activation.
The latter is composed of a \textit{warm} module interfacing with the \gls{daq} and a \textit{cold} electronics circuit on the \gls{pdu} motherboard. 
The control through the steering system has several functions: turning on and off individual channels, turning on and off the \gls{lv} and \gls{hv} of the \(16\) Tiles independently and turning on and off a microcontroller needed to send a \gls{pdu} unique identifier to the \gls{daq}. A unique bias voltage per \gls{pdu}, hence four analogue channels, is present.

Using the CERN WinCC-OA Supervisory Control and Data Acquisition (SCADA) framework ensures stable, real-time monitoring and data exchange to and from the \gls{daq} for detector and status validation~\cite{winccoa}. Information exchange and direct device access enhance redundancy and reliability.

A Detector Safety System (DSS) based on a dedicated Programmable Logic Controller (PLC) safeguards equipment and personnel by interfacing with the cryogenic and \gls{tpc} subsystems. It connects directly to \gls{tpc} hardware and \gls{pdu} power supplies through interlocks managed over an Uninterruptible Power Supply (UPS)-backed network, with additional interlocks securing the cryogenic system.


\section{Quadrant Test}
\label{sec:quadrant}

\begin{figure}[tpb]
    \centering
    \includegraphics[width=0.6\linewidth]{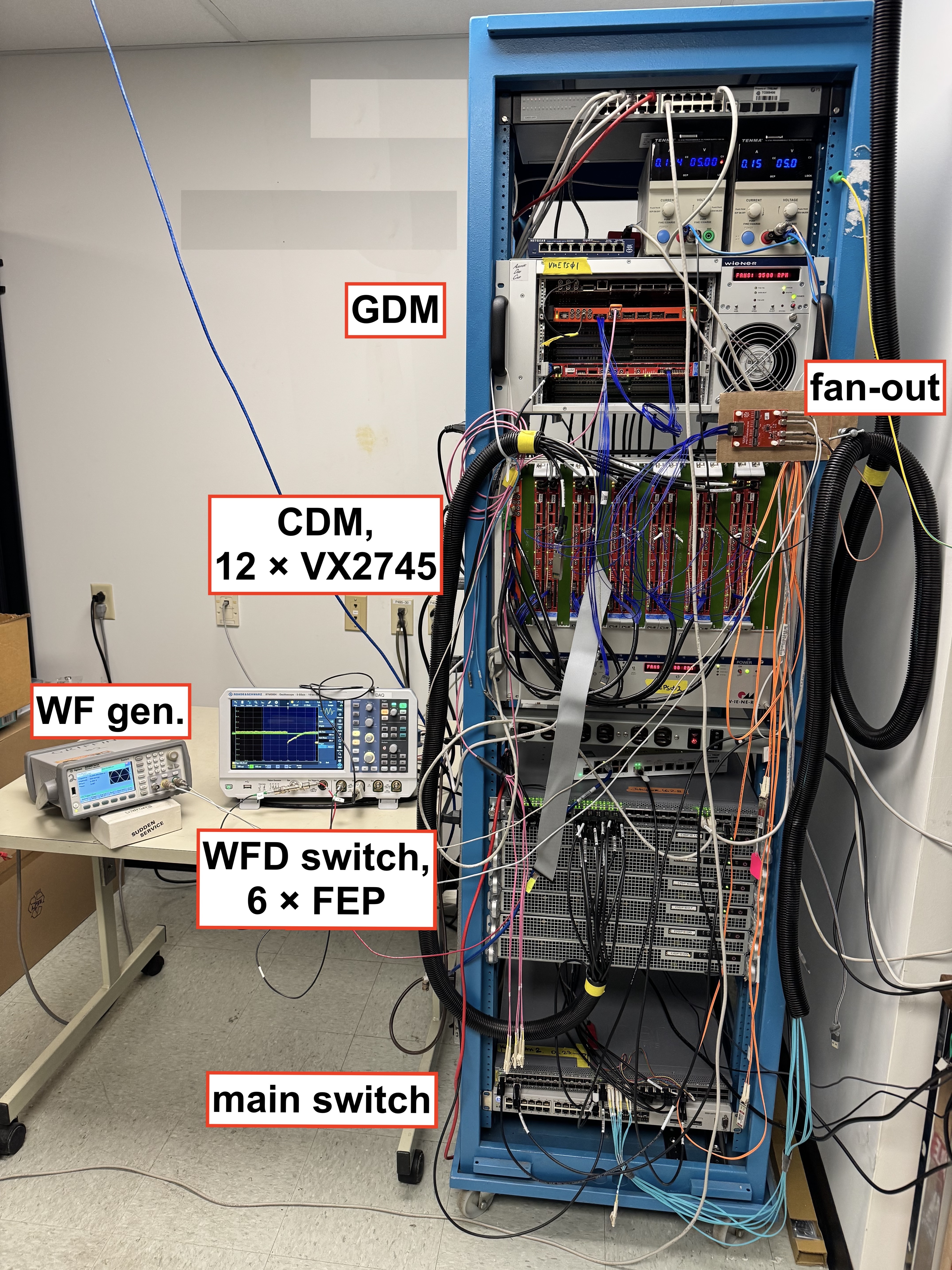}
    \caption{The picture shows the Quadrant setup built and tested in TRIUMF, Canada. It corresponds to one of the final system's four identical racks, including \(12\) waveform digitisers (\wfds) and \(6\) \glspl{fep}. Not shown are the \glspl{tsp} located in a separate rack and the \gls{midas} Supervisor.}
    \label{fig:quadrant}
\end{figure}

A "Quadrant", i.e. 1/4 of the \gls{ds20k} \gls{daq} system, has been realised at the TRIUMF Laboratory, Canada, for development and testing (see \autoref{fig:quadrant}). It was composed of one \gls{gdm} and one \gls{cdm} board, \(12\) \digmodel\ boards, one \SI{10} GbE network switch, \(6\) \glspl{fep}, optical link connection to \(5\) \glspl{tsp}, external clock interface and an analogue fan-out connected to the waveform generator for synchronisation tests. A \SI{64}{\giga \byte} memory SuperMicro AS-2015A-TR machine has been chosen to host the \gls{midas} supervisor.

\begin{figure*}
    \centering
    \includegraphics[width=0.75\linewidth]{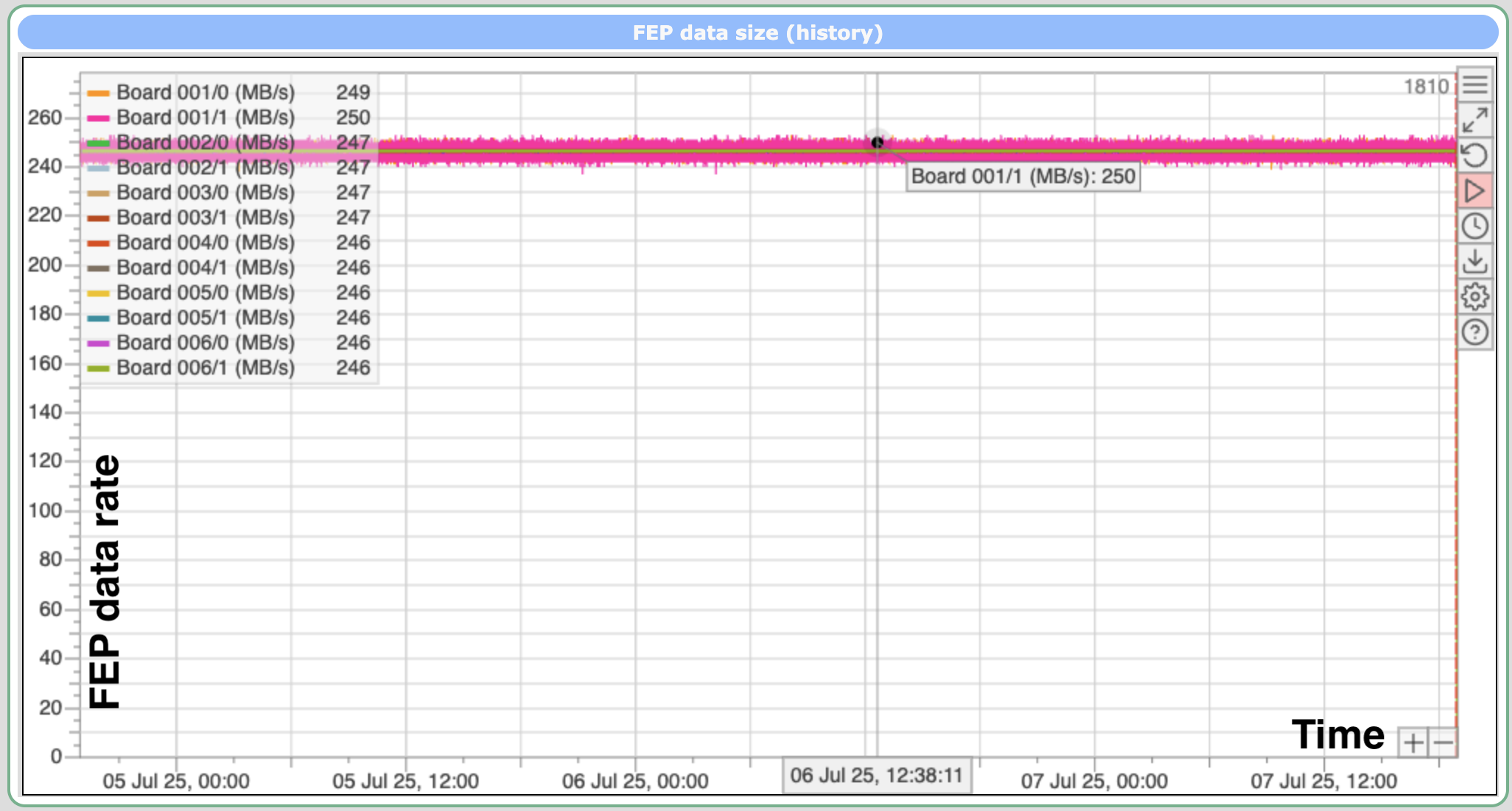}
    \caption{Plot presents the data rate in MB/s from 12 digitiser boards to the \gls{fep} machines over a period of 2.5 days. The rate was stable for all boards at \SI{250} MB/s.}
    \label{fig:fep_rate}
\end{figure*}

The Quadrant setup has been stress-tested by using a 2 kHz periodic trigger simultaneously on all the \(768\) channels from the \(12\) digitisers. This configuration corresponds to the worst-case scenario when all the channels are triggering at the same time. The length of the acquired waveforms was set to \SI{8}{\micro \second}. A sustainable data rate of \SI{250} MB/s has been measured on each digitiser board without any lagging and is limited by the \gls{fep} data handling performance, while the connection between the digitiser board and the \gls{fep} allows for higher data rate transfer of \SI{10} GbE, ensuring a safety margin.  The measured rate for a single \gls{fep} is presented in ~\autoref{fig:fep_rate}. The \gls{fep} processor usage is below 60\% for any thread. The data rate from the \gls{fep} to the \gls{tsp} during this test was \SI{8} MB/s. The system demonstrated stable operation over long operation (300 h). Finally, the full \gls{ts} architecture has been implemented and tested with part of the Quadrant setup.

\section{Conclusions}
One of the challenges of \gls{ds20k} experiment is the development of a robust \gls{daq} system enabling continuous signal acquisition from its 2720 channels. The successful implementation of the \gls{daq} system architecture presented here demonstrates that such a system can be built largely from commercial components allowing for easy scalability. The CAEN \digmodel\ digitisers provide high channel density and flexibility thanks to the programmable OpenFPGA. Multi-board synchronisation of 48 modules with custom Data Manager boards ensuring sub-nanosecond synchronisation across all digitisers through a rubidium standard disciplined by \gls{lngs} signal enables coherent sampling and accurate time segmentation for event reconstruction and absolute time information essential for the supernova events.

The distributed processing based on Front End Processors and a scalable farm of Time Slice Processors allows for online hit extraction, data reduction, event classification, and possible additional algorithms without impacting the data flow.

The successful commissioning of the Quadrant system in TRIUMF representing one quarter of the final \gls{daq} confirms the feasibility of the full design. Sustained operation at \SI{250} MB/s per digitiser with simultaneous trigger demonstrates that the system satisfies the throughput and stability required for physics data taking. The data synchronisation test showed coherent timing of the data acquisition with an intra-channel spread well below the sampling period.

\printglossary[type=\acronymtype, title=List of Acronyms]

\appendix

\acknowledgments

We would like to thank the other scientists and technical staff for their essential contributions throughout the course of this work, in particular Ian Johnson (jTechnologies) for the development of the digitiser firmware, Peter Margetak (TRIUMF) for the design and production of custom boards, and Samuel de Jong (University of Victoria) for his contributions to the FPGA firmware.
We also acknowledge the support and assistance provided by the staff at CAEN.

This work was supported by the U.S. National Science Foundation
(NSF) through Grants No. PHY-0919363, PHY-1004054, PHY-1004072,
PHY-1242585, PHY-1314483, PHY-1314507, PHY-1622337, PHY-1812482,
PHY-1812547, PHY-2310091, PHY-2310046, associated collaborative
grants No. PHY-1211308, PHY-1314501, PHY-1455351 and PHY-1606912, as
well as Major Research Instrumentation Grant No. MRI-1429544.
Additional support was provided by the Pacific Northwest National
Laboratory, operated by Battelle for the U.S. Department of Energy
under Contract No. DE-AC05-76RL01830.

Support was provided by the Istituto Nazionale di Fisica Nucleare
(INFN), through grants from the Italian Ministero dell’Istruzione,
Università e Ricerca, including Progetto Premiale 2013 and
Commissione Scientifica Nazionale II, as well as by the PRIN2020
project of the Italian Ministry of Research (MUR) (Grant No. PRIN
20208XN9TZ).

This work was supported by Canada Foundation for Innovation (CFI), the Natural Sciences and Engineering
Research Council of Canada, SNOLAB, and the Arthur B. McDonald
Canadian Astroparticle Physics Research Institute.
Support was received from the French government from LabEx
UnivEarthS (ANR-10-LABX-0023 and ANR-18-IDEX-0001). Additional
support was received from the IN2P3-COPIN consortium (Grant No.
20-152).

This work was supported by the Chinese Academy of Sciences
(113111KYSB20210030) and the National Natural Science Foundation of
China (12020101004).
Support was provided by the São Paulo Research Foundation (FAPESP)
under Grant No. 2021/11489-7 and by the National Council for
Scientific and Technological Development (CNPq).
Support is acknowledged from the Deutsche Forschungsgemeinschaft
(DFG, German Research Foundation) under Germany’s Excellence
Strategy — EXC 2121: Quantum Universe — 390833306.

The authors acknowledge support from the Spanish Ministry of Science
and Innovation (MICINN) through Grants PID2022-138357NB-C22 and
PID2022-138357NB-C21 and the Atracción de Talento Grant 2018-T2/
TIC-10494.

This work was supported by the Polish National Science Centre (NCN)
through Grants No. UMO-2022/47/B/ST2/02015 and UMO-2023/51/B/
ST2/02099, by the Polish Ministry of Science and Higher Education
(MNiSW, Grant No. 6811/IA/SP/2018).
This work was supported by the FNP IRA programmes: AstroCeNT
(MAB/2018/7), funded from the ERDF, and Astrocent
(FENG.02.01-IP.05-A015/25) co-financed by the European Union under
FENG 2021–2027; and Teaming for Excellence grant Astrocent Plus
(101137080)
funded by the European Union with complementary national funding from
the MNiSW (MNiSW/2025/DIR/811).

This project received funding from the European Union’s Horizon 2020
research and innovation programme under Grant Agreement No. 952480
(DarkWave).

Support was provided by the Science and Technology Facilities
Council, part of United Kingdom Research and Innovation, and by The
Royal Society.



\bibliographystyle{JHEP}
\bibliography{bibliography.bib}


\end{document}